%% file: paper.tex
\documentclass[11pt]{article}
\input{preamble}
\hypersetup{pdftitle={Phase-resolved spin-orbit exciton transfer with dressed quantum encodings}}
\title{\Large\bfseries Phase-resolved spin--orbit exciton transfer with dressed quantum encodings}
\author{Y\"ucel Y\i ld\i r\i m$^{1}$, \"Ozg\"ur Polat$^{2}$, Cengiz \c{S}en$^{3}$\\[5pt]
\parbox{.94\textwidth}{\centering\small $^1$Department of Mechanical Engineering, Faculty of Engineering and Natural Sciences, Istanbul Bilgi University, Ey\"upsultan 34060, Istanbul, T\"urkiye\\[3pt]$^2$Department of Physics and Astronomy, California State Polytechnic University, 3801 West Temple Ave, Pomona, CA 91768, USA\\[3pt]$^3$Department of Physics, Lamar University, Beaumont, TX 77710, USA\\[3pt]\texttt{yucel.yildirim@bilgi.edu.tr}}}
\date{}
\begin{document}
\maketitle

\begin{abstract}\noindent
A quantum simulation of a material runs on a register much smaller than the electronic problem behind it, and the map between the two decides what a measured correlation means. We study this map for spin--orbit exciton transfer across an Ir--Ir bond of a LiIrF$_6$-derived model. Exact references show that a bare product-state encoding has an amplitude floor above a 0.02 state tolerance at sampled times, whereas a polar-dressed encoding lowers the ten-window projected error from 0.320 to $3.54\times10^{-4}$. Three-CZ axial and two-CZ transverse circuits, run on IBM Kingston in eight acquisitions with 1.6 million shots, resolve population transfer together with both quadratures of the intersite coherence. Raw half-transfer Bell-projector scores reach 0.935 in two independent acquisitions, and on a second bond a coherent arm with lower bound 0.871 is separated from a matched dephased control. Across 2,520 model points, adaptive spectral encodings meet fixed accuracy targets with explicit dimension costs.
\end{abstract}

\section*{Introduction}
A quantum processor that simulates a material works with a small model of it. The model is chosen before any circuit is written, and the choice already fixes what the measured numbers mean. A state called an exciton on the logical register may stand for a bare local excitation, a dressed combination of local configurations or a compressed eigenstate. These options can share an effective spectrum and still prescribe different initial states and different observable maps. For near-term devices, where every qubit and every entangling gate is expensive~\cite{Bauer2020,Daley2022,Kim2023Utility}, the reduction is unavoidable; its consequences for a measured population or coherence deserve the same accounting as gate errors.

Spin--orbit-coupled $d^4$ insulators give the question a concrete physical form. Strong spin--orbit coupling in a $5d^4$ ion produces a nonmagnetic $J=0$ singlet and low-lying triplet excitations~\cite{JackeliKhaliullin2009,WitczakKrempa2014,Takayama2021}. Intersite hopping lets these excitations propagate, while pair creation and anisotropic terms compete with the hopping and can drive excitonic magnetism~\cite{Khaliullin2013,Meetei2015,Jain2017,Kim2019}. Propagating spin--orbit excitons have been resolved by resonant inelastic x-ray scattering in pentavalent iridates such as Ba$_2$YIrO$_6$~\cite{Kusch2018,Nag2018}. The fluoride LiIrF$_6$ belongs to the same $5d^4$ family, whose specified model has Ir--Ir couplings below one millielectronvolt; its local fine structure is analysed in two companion studies~\cite{YildirimMaterial1,YildirimMaterial2}.

Selecting one triplet component on each ion reduces a bond to a two-site transfer problem. The remaining components still act virtually, so the bare product-state subspace is not invariant under the bond Hamiltonian. A circuit that implements the projected Hamiltonian perfectly then accumulates a phase error or loses amplitude. Schrieffer--Wolff theory and canonical effective Hamiltonians describe the virtual dressing~\cite{Schrieffer1966,desCloizeaux1960,Bravyi2011}. An experiment also needs a consistent rule for the initial state and the measured operators; rotated-basis measurements based on this principle have been demonstrated for Fermi--Hubbard simulators~\cite{Kale2022Rotated}. Here we ask how the same principle enters the error budget of a material-derived multilevel bond, and how it shows up in phase-sensitive finite-shot tests.

Phase resolution is the second ingredient. A transferred population can also arise from an incoherent mixture. Measuring both quadratures of the intersite coherence tests the prepared state much more sharply and allows a direct comparison with a deliberately dephased control. Quantum hardware has already reconstructed dynamical response functions and simulated molecular and material dynamics~\cite{Chiesa2019,Francis2020,Maskara2025}; the interpretation of such data depends on readout calibration and on whether the ensemble has been conditioned on a symmetry sector~\cite{Bonet2018,McArdle2019,Bravyi2021Readout,Nation2021}. Observable-specific analyses of simulation error point in the same direction: the operator that is measured controls the error that matters~\cite{Childs2021,Granet2025Dilution,Yu2025Observable}.

We combine these elements in one study. The material input is a specified Wannier--Kanamori model of a LiIrF$_6$ bond, reconstructed independently in the 495-dimensional two-ion space. Two dressed four-state encodings lead to an axial experiment that tracks the site of one exciton and a transverse experiment that also tracks its flavour. Eight acquisitions on IBM Kingston include independent repetitions, a second bond and new fixed-witness and dephasing tests. On the theory side, a singular-value identity gives the best full-state accuracy available inside a bare encoding, and a matched preparation-and-observable bound explains why a population can stay accurate when that state accuracy is out of reach. Tests on 21 connected bonds and a 2,520-point model class then show when four logical states suffice and supply sufficient larger representations otherwise.

\section*{Results}
\subsection*{A bond model with two encodings}
\begin{figure}[t]\centering
\includegraphics[width=\textwidth]{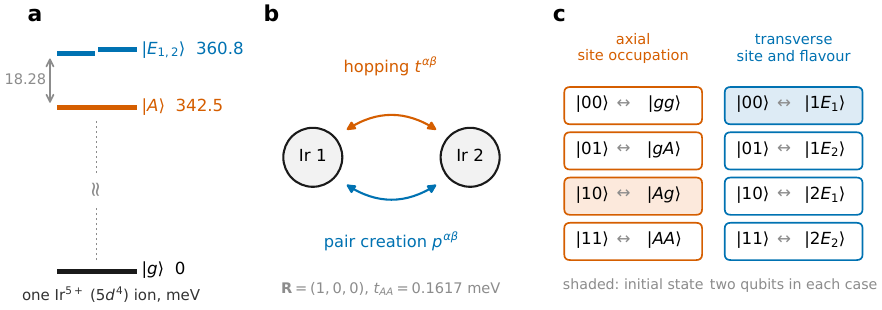}
\caption{\textbf{Bond model and logical encodings.} \textbf{a}, Local levels of one Ir$^{5+}$ ion in the specified model (not to scale); the axial state lies 18.28~meV below the transverse pair. \textbf{b}, The two-ion bond couples the ions through exciton hopping $t^{\alpha\beta}$ and pair creation $p^{\alpha\beta}$; the primary bond has bare axial hopping $t_{AA}=0.1617$~meV. \textbf{c}, Axial encoding by site occupation and transverse encoding by site and flavour. Shaded boxes mark the initial states used in the experiments.}\label{fig:model}
\end{figure}

Each ion carries a singlet $|g\rangle$ and three triplet-derived states $|A\rangle$, $|E_1\rangle$ and $|E_2\rangle$ (Fig.~\ref{fig:model}a). At $U=3$~eV and $J_H=0.4$~eV their excitation energies are 342.486927, 360.756199 and 360.768994~meV, so the axial state lies about 18.28~meV below the transverse pair. With $T^\dagger_{i\alpha}=|\alpha\rangle_i\langle g|$, the two-site Hamiltonian reads
\begin{equation}
H=\sum_{i\alpha}\epsilon_\alpha T^\dagger_{i\alpha}T_{i\alpha}
+\sum_{\alpha\beta}\left(t^{\alpha\beta}T^\dagger_{1\alpha}T_{2\beta}
+p^{\alpha\beta}T^\dagger_{1\alpha}T^\dagger_{2\beta}+\mathrm{h.c.}\right)+H_{\rm rest},
\label{eq:bond}
\end{equation}
where $H_{\rm rest}$ collects the remaining connected terms of the $16\times16$ bond matrix. All of them are kept in the reference calculations.

For the primary bond, $\mathbf R=(1,0,0)$, the bare axial hopping is $t_{AA}=0.161659831$~meV and the corresponding transfer time is $t_A^{(0)}=\pi\hbar/(2|t_{AA}|)=6395.63$~fs. These times parameterize evolution under the material model and are unrelated to processor gate durations. The axial encoding uses site occupation, $|q_1q_0\rangle=|gg\rangle,|gA\rangle,|Ag\rangle,|AA\rangle$, and starts in $|Ag\rangle=|10\rangle$. The transverse encoding describes a single $E$ excitation: one qubit records its site and the other its $E_1/E_2$ flavour (Fig.~\ref{fig:model}c). The two registers have the same size and answer different physical questions.

The native two-ion reference has eight electrons in twelve spinors, giving $\binom{12}{8}=495$ determinants. Two independent reconstructions agree to $1.56\times10^{-10}$~meV in the connected bond matrix after conversion to a common Fock basis. Reduced targets obtained directly from this reference and from the 16-state model differ by at most $2.80\times10^{-5}$~meV per entry in the axial channel and $4.74\times10^{-10}$~meV in the transverse channel, and their sampled projected-dynamics discrepancies stay below $8.05\times10^{-4}$. These checks fix the numerical operator that the experiments test.

The excitation centroid is $\Delta=(\epsilon_A+\epsilon_{E_1}+\epsilon_{E_2})/3=354.670707$~meV, with signed pair-minus-axial splitting 18.275670~meV and pair defect 0.012795~meV; each $\epsilon$ is measured from the same local ground state. These are parameters of the native Wannier--Kanamori model. The electronic parent is constructed classically, its dressed four-state dynamics are compiled onto two processor qubits, and material observables are reconstructed using the same dressing map. The final correlated-cluster calculation of the companion material study is a separate operator. At fixed local electron number, changing $U$ adds a common Kanamori shift, whereas virtual charge excitations entering exchange can change (Supplementary Note~\ref{SI-sec:model}); fitting local excitation energies alone therefore does not determine the bond interaction.

\subsection*{The bare encoding sets an accuracy floor}
\begin{figure}[t]\centering
\includegraphics[width=\textwidth]{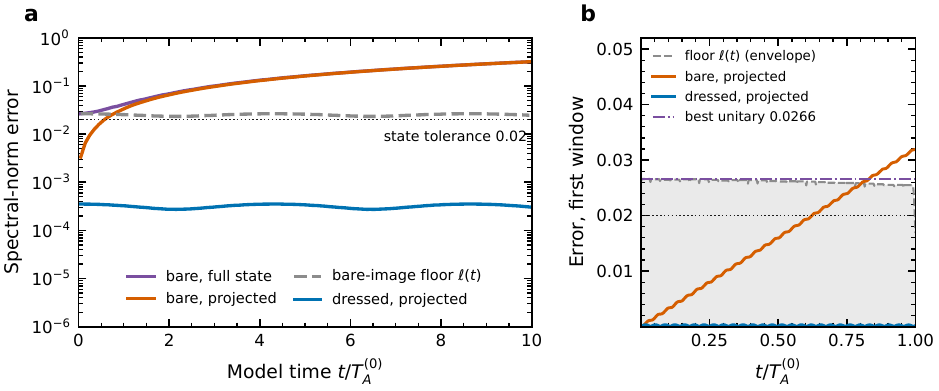}
\caption{\textbf{Representation errors of the primary axial bond.} \textbf{a}, Spectral-norm errors over ten bare transfer windows for the bare encoding (full state and projected propagator), the bare-image floor $\ell(t)$ of Eq.~\eqref{eq:floor} and the dressed encoding. Curves are maxima over 100 time bins of a 4,001-point exact grid, which summarizes the fast coherent beating while retaining the sampled maxima. The dotted line marks the 0.02 state tolerance. \textbf{b}, First transfer window. The floor $\ell(t)$ (envelope of the fast beating) and the optimal bare unitary of Eq.~\eqref{eq:minimax} both exceed the tolerance, while the dressed projected error stays at the $10^{-4}$ level.}\label{fig:reduction}
\end{figure}

Let $W$ be the isometry from four logical states into the bare product-state subspace, $P=WW^\dagger$, and $U(t)=\exp(-iHt/\hbar)$. For every logical unitary $V$, orthogonal projection gives
\begin{equation}
\|U(t)W-WV\|_2\ge\ell(t),\qquad \ell(t)=\|(1-P)U(t)W\|_2 .
\label{eq:floor}
\end{equation}
The leakage $\ell$ is an amplitude and $\ell(0)=0$; a positive floor is asserted only at the specified nonzero times. A probability of $10^{-3}$ outside the subspace already corresponds to an amplitude of 0.0316, above a state-amplitude tolerance of 0.02, and no circuit whose output stays inside the bare image can remove it. The best logical unitary follows from the polar factor of $W^\dagger UW$, and its exact spectral-norm distance is
\begin{equation}
\min_{V\ \mathrm{unitary}}\|UW-WV\|_2=\sqrt{2\left[1-\sigma_{\min}(W^\dagger UW)\right]}
\label{eq:minimax}
\end{equation}
(Supplementary Note~\ref{SI-sec:proofs}). In the first transfer window of the primary axial bond the sampled leakage reaches 0.0265922 and the optimal distance is 0.0265945 (Fig.~\ref{fig:reduction}b). A 0.02 state tolerance is therefore out of reach for any bare four-state description at those times.

Dressing changes the image of the encoding. Let $X$ hold the four selected eigenvectors of $H$ with eigenvalues $E$, and write $W^\dagger X=U_A\Sigma V_A^\dagger$. The polar rotation $R=U_AV_A^\dagger$ defines
\begin{equation}
W_d=XR^\dagger,\qquad H_d=RER^\dagger,\qquad UW_d=W_de^{-iH_dt/\hbar}.
\label{eq:dressed}
\end{equation}
Among all orthonormal bases of the invariant subspace, $W_d$ is the one closest to $W$ in Frobenius norm~\cite{Lowdin1950,Schonemann1966}, and $H_d$ is the corresponding canonical effective Hamiltonian~\cite{desCloizeaux1960}. Preparations and observables transform with the same isometry. On the processor $e^{-iH_dt/\hbar}$ acts on the logical register, and $W_d$ tells us what the register means in the material model. For the primary axial bond a second-order Schrieffer--Wolff construction agrees with $H_d$ within $1.93\times10^{-6}$~meV per entry. Dressing shifts the one-exciton diagonal elements by about $-0.0020$~meV and raises the hopping from 0.161660 to 0.162959~meV. The change is below one percent, yet the phase it accumulates is easily resolved over several transfer windows: after ten windows the bare projected error has grown to 0.320, while the dressed error stays at $3.54\times10^{-4}$ (Fig.~\ref{fig:reduction}a and Table~\ref{tab:reduction}).

\begin{table}[t]\centering\small
\caption{\textbf{Spectral-norm errors of the primary axial reduction.} One and ten transfer windows are sampled with 401 and 4,001 time points. Population entries are total-variation errors for the $|Ag\rangle$ input.}\label{tab:reduction}
\begin{tabular}{lcc}\toprule
Quantity & One window & Ten windows\\\midrule
Bare-image leakage amplitude $\ell$ & 0.026592 & 0.026600\\
Bare full-state isometry error & 0.040368 & 0.320297\\
Bare projected-propagator error & 0.031945 & 0.319635\\
Dressed projected-propagator error & $3.537\times10^{-4}$ & $3.538\times10^{-4}$\\
Population error, bare & 0.00741 & 0.1194\\
Population error, dressed & 0.00042 & 0.00042\\\bottomrule
\end{tabular}
\end{table}

\subsection*{Matched preparations and observables}
The dressed projected error in Table~\ref{tab:reduction} still refers to a bare preparation, which is what a simple experiment would load. For such inputs a matched pair of preparation and observable maps gives an explicit budget. Write $W=JA+R_\perp$, where $J$ is an invariant dressed isometry, $A=J^\dagger W$ and $R_\perp=(1-JJ^\dagger)W$. With $\alpha=\|A\|_2$, $\rho=\|R_\perp\|_2$ and $Q=1-JJ^\dagger$, the matched dressed prediction $F_d=A^\dagger V^\dagger J^\dagger OJVA$ obeys
\begin{equation}
\|W^\dagger U^\dagger OUW-F_d\|_2\le2\alpha\rho\|J^\dagger OQ\|_2+\rho^2\|QOQ\|_2 .
\label{eq:observable}
\end{equation}
An observable whose retained--excluded cross block vanishes therefore carries a representation error quadratic in $\rho$, although the state error is linear in $\rho$. This is why a population can remain accurate where an all-input state guarantee fails. Supplementary Note~\ref{SI-sec:proofs} gives the proof, its extension to a finite residual and the bound for conditioned observables. Numerical checks on 117 native, rescaled and synthetic cases, with seven observables and 61 time points each, find a largest excess over the bound of $6.93\times10^{-15}$; a second set of 126 combinations of bond, encoded branch and symmetric assignment channel also stays within its bounds.

The observable map completes the encoding. For a logical density matrix $\rho_L$, a model operator $O$ has the prediction
\begin{equation}
\langle O\rangle_d=\Tr\!\left[\rho_L\,W_d^\dagger OW_d\right],\qquad W_d^\dagger OW_d=\sum_{\mu,\nu=0}^{3}o_{\mu\nu}\,\sigma_\mu\otimes\sigma_\nu ,
\label{eq:readoutmap}
\end{equation}
and the hardware measures the Pauli terms selected by each protocol. Logical occupations and Bell projectors are direct register observables; a bare material occupation uses the transformed coefficients $o_{\mu\nu}$. Because dressing mixes the site factors, entanglement between the two processor qubits refers to the logical register, and its material meaning follows from $W_d$.

\subsection*{Two- and three-CZ circuits}
After removing a scalar shift and ordering each Pauli string as $q_1q_0$, the dressed targets read, in meV,
\begin{align}
H_A={}&-171.2232(Z_1+Z_0)+0.19294\,Y_1Y_0-0.02998\,X_1X_0-0.02721\,Z_1Z_0\nonumber\\
&+0.00514(X_1+X_0)+0.00137(X_1Z_0+Z_1X_0),\\
H_E={}&X_1\otimes(0.12120\,I+0.08095\,X_0-0.02818\,Z_0)+0.03684\,X_0+0.03227\,Z_0 .
\end{align}
The rounded coefficients are for interpretation only; circuits and checks use the full stored matrices. The $(XX+YY)/2$ term gives the dressed hopping of 0.162959~meV. The $(XX-YY)/2$ coefficient, about $-0.2229$~meV, couples $gg$ to $AA$ across a detuning of about 684.9~meV; it preserves parity, whereas the weak $X$ and $XZ$ terms change the exciton number by one and mix the parity sectors.

\begin{figure}[t]\centering
{\small\textbf{a}}\quad
\begin{minipage}{.46\textwidth}\centering\resizebox{\linewidth}{!}{\input{circuits/xa_logical.tex}}\end{minipage}\hfill
\begin{minipage}{.46\textwidth}\centering\resizebox{\linewidth}{!}{\input{circuits/xe_logical.tex}}\end{minipage}\\[7pt]
{\small\textbf{b}}\\[1pt]\resizebox{\textwidth}{!}{\input{circuits/XA_original_half_XX_tw0.tex}}\\[7pt]
{\small\textbf{c}}\\[1pt]\resizebox{.87\textwidth}{!}{\input{circuits/XE_original_half_XX_tw0.tex}}
\caption{\textbf{From effective Hamiltonian to native gates.} \textbf{a}, Logical axial (left) and transverse (right) programs with Pauli-basis readout. \textbf{b}, Executed axial circuit at half transfer, $XX$ setting: three CZ gates. \textbf{c}, Executed transverse circuit at half transfer, $XX$ setting: two CZ gates. Native circuits are drawn from the executed payloads, including preparation and readout rotations. Displayed angles are rounded for legibility; execution used the stored payload values. Logical wires $(q_0,q_1)$ map to physical qubits $(21,36)$ of IBM Kingston. Pauli randomizations preserve the ideal output distribution to $1.7\times10^{-15}$.}\label{fig:circuits}
\end{figure}
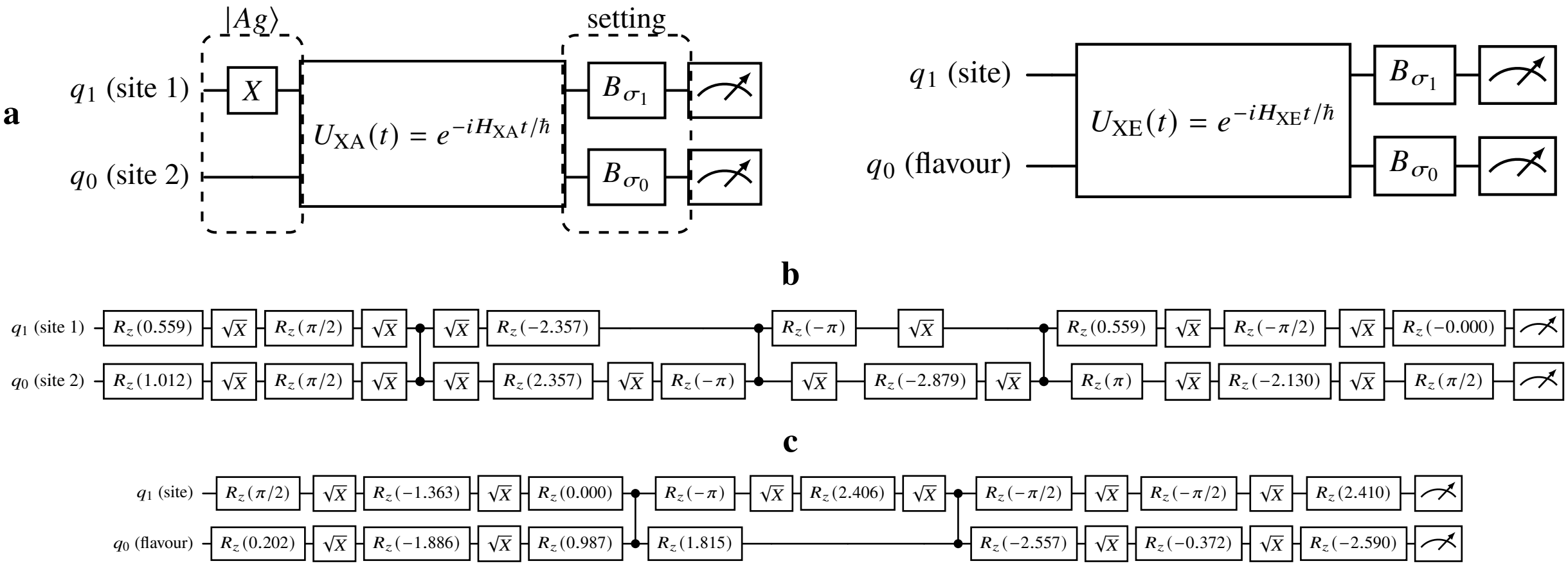

The dressed axial and transverse transfer windows are $T_A=6344.63$~fs and $T_E=5684.79$~fs. At each requested time we compile the full logical unitary by direct synthesis. Its Cartan form,
\begin{equation}
U=(k_1\otimes k_2)\,e^{i(aXX+bYY+cZZ)}\,(k_3\otimes k_4),
\end{equation}
separates local rotations from the nonlocal content~\cite{Khaneja2001,Kraus2001,Zhang2003}. The actual axial propagators at the four executed nonzero times have $c\ne0$ in the Weyl chamber (Table~\ref{tab:weyl-main}) and require three CZ gates for exact synthesis up to local rotations~\cite{Vatan2004,VidalDawson2004,Shende2004}. This follows from their nonlocal invariants, rather than from the presence of a particular Hamiltonian term alone. For the transverse Hamiltonian, $[H_E,X_1]=0$, and the $X_1$ eigenbasis resolves its propagator as $|+\rangle\langle+|\otimes u_++|-\rangle\langle-|\otimes u_-$. Factoring out one local unitary gives a controlled unitary, with Weyl coordinates $(a,0,0)$. At the sampled angles it compiles exactly with two CZ gates (Fig.~\ref{fig:circuits}). This is a generic controlled-unitary count: at special angles a locally equivalent CNOT needs only one entangler, and a local operation, including zero-time evolution, needs none. The distinction concerns full logical propagators at a common synthesis tolerance; a restricted input or observable can admit a cheaper circuit.

\begin{table}[h]\centering\small
\caption{Weyl coordinates of the executed nonzero-time targets (radians), for $\exp[i(aXX+bYY+cZZ)]$. Local operations are removed and the chamber uses $\pi/4\ge a\ge b\ge |c|$. Entangler counts refer to these four time points; zero time is local.}\label{tab:weyl-main}
\begin{tabular}{crrrrrr}\toprule
&\multicolumn{3}{c}{Axial, three CZ}&\multicolumn{3}{c}{Transverse, two CZ}\\\cmidrule(lr){2-4}\cmidrule(lr){5-7}
$t/T$&$a$&$b$&$c$&$a$&$b$&$c$\\\midrule
0.25&0.1966&0.1961&0.0656&0.1848&0.0000&0.0000\\
0.50&0.3930&0.3924&0.1311&0.3679&0.0000&0.0000\\
0.75&0.5891&0.5890&0.1967&0.5473&0.0000&0.0000\\
1.00&0.7851&0.7851&$-0.2623$&0.7194&0.0000&0.0000\\\bottomrule
\end{tabular}
\end{table}

\subsection*{Coherent axial transfer in two acquisitions}
Axial measurements use the settings $ZZ$, $XX$, $YY$, $XY$ and $YX$. With the site order fixed above, the intersite coherence is
\begin{equation}
n_{12}=\frac{\langle XX\rangle+\langle YY\rangle}{4}+\frac{i}{4}\left(\langle XY\rangle-\langle YX\rangle\right).
\label{eq:coherence}
\end{equation}
A positive hopping acting on $|Ag\rangle$ produces a negative imaginary coherence near half transfer, a sign that is meaningful relative to this phase convention. Pair creation changes the exciton number by two and preserves parity, while weak single-$X$ and $XZ$ terms in the target mix the even and odd sectors. Raw observables and results conditioned on measured odd parity are therefore kept separate throughout.

\begin{figure}[t]\centering
\includegraphics[width=\textwidth]{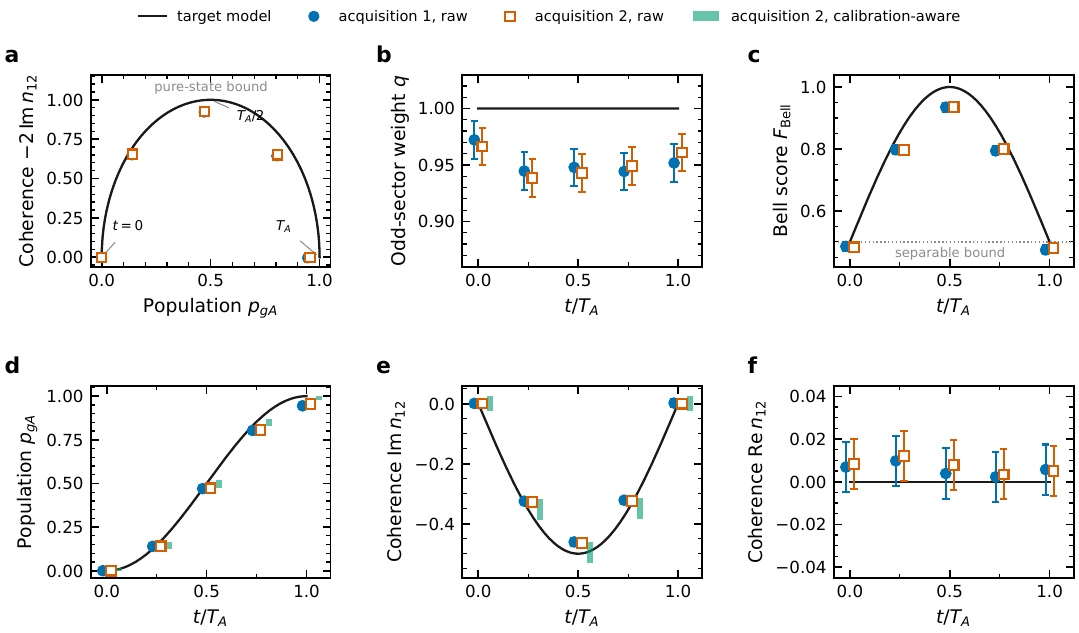}
\caption{\textbf{Coherent axial transfer.} \textbf{a}, Phase-space trajectory of $-2\Im n_{12}$ against transferred population $p_{gA}$; the dashed curve is the pure one-exciton bound. \textbf{b}, Raw odd-sector retention $q=p_{01}+p_{10}$. \textbf{c}, Raw fixed Bell-projector score, with separable bound $1/2$. \textbf{d}--\textbf{f}, Transferred population and imaginary and real coherence; the real-part scale is enlarged. Filled circles and open squares are raw estimates from two acquisitions, with simultaneous 95\% intervals over 210 linear quantities. Horizontal and vertical intervals in \textbf{a} use that same family. Green bars are calibration-aware, parity-conditioned intervals for the second acquisition (preparation allowance 0.002). Black curves are the dressed target model. Time is measured in units of $T_A$.}\label{fig:axial}
\end{figure}

The four primary acquisitions contain 1,318,912 shots. Both axial acquisitions resolve the expected coherence (Fig.~\ref{fig:axial}). In the first, the midpoint value is $\Im n_{12}=-0.46091$ with a magnitude lower bound of 0.44613, and the raw transferred population at the final time is 0.94489. The second acquisition, taken after a later device calibration, reproduces the shape of every trajectory. The population--coherence trajectory in Fig.~\ref{fig:axial}a makes the phase information explicit. A pure one-exciton state at the transfer phase traces $2[p_{gA}(1-p_{gA})]^{1/2}$, whereas a mixture diagonal in the site basis has zero intersite coherence. The measured trajectory follows the coherent arc, with raw odd-sector retention shown separately in Fig.~\ref{fig:axial}b. This comparison uses trusted Pauli measurements of a common state; it is not inferred from population transfer alone. A compact phase-sensitive summary is the fixed Bell projector of ideal half transfer,
\begin{equation}
F_{\rm Bell}=\frac{1-\langle ZZ\rangle-\langle XY\rangle+\langle YX\rangle}{4}.
\label{eq:bell}
\end{equation}
Its raw values are 0.934845 and 0.935059, with simultaneous shot-only intervals $[0.920464,0.949226]$ and $[0.920677,0.949440]$, obtained without assignment inversion or output conditioning. For trusted local Pauli measurements of a common prepared state, $F_{\rm Bell}$ is a projector overlap and exceeds the separable bound of $1/2$ by a wide margin~\cite{GuhneToth2009}.

Raw and calibrated analyses address different accuracy questions. The raw simultaneous intervals resolve both the magnitude and the sign of the midpoint coherence in the two acquisitions, while the stricter requirement that every raw population and coherence interval lie within $\pm0.05$ of the target is exceeded at some points in both. For the second acquisition we also construct calibration-aware outer intervals, assuming a common assignment channel for calibration and signal and allowing a calibration-preparation error of 0.002 in total variation. All four criteria then hold within $\pm0.05$; the midpoint interval for $\Im n_{12}$ is $[-0.53147,-0.46062]$ (Supplementary Note~\ref{SI-sec:stats}). Table~\ref{tab:outcomes} lists the endpoints used to judge every acquisition.

\begin{table}[t]\centering\small
\caption{\textbf{Experimental endpoints and outcomes.} Raw, corrected and conditioned quantities are separate endpoints. The criteria of the last three experiments were fixed before their counts were collected.}\label{tab:outcomes}
\begin{tabular}{>{\raggedright\arraybackslash}p{.24\textwidth}>{\raggedright\arraybackslash}p{.27\textwidth}>{\raggedright\arraybackslash}p{.40\textwidth}}\toprule
Experiment & Endpoint & Result\\\midrule
Primary axial, two acquisitions & Raw midpoint coherence magnitude and sign; raw $\pm0.05$ envelope & Magnitude and sign resolved in both; envelope exceeded at some points in both\\
Axial repeat & Calibration-aware conditional accuracy & All four criteria met at preparation allowance 0.002\\
Primary transverse, two acquisitions & Raw site population, conditional flavour, midpoint coherence & All criteria met in both\\
Second-bond populations, two acquisitions & Four-state accuracy and bare/dressed contrast & First: all four criteria met. Repeat: contrast met; two dressed TV bounds at 0.041 and 0.044 against 0.04\\
Fixed transverse witness & Linear-coherence lower bound & Normalized lower bound 0.852355 against 0.8\\
Second-bond phase test & Coherent and dephased arms & All six criteria met; coherent Bell lower bound 0.870609\\\bottomrule
\end{tabular}
\end{table}

\subsection*{Complex coherence in the transverse experiment}
\begin{figure}[t]\centering
\includegraphics[width=\textwidth]{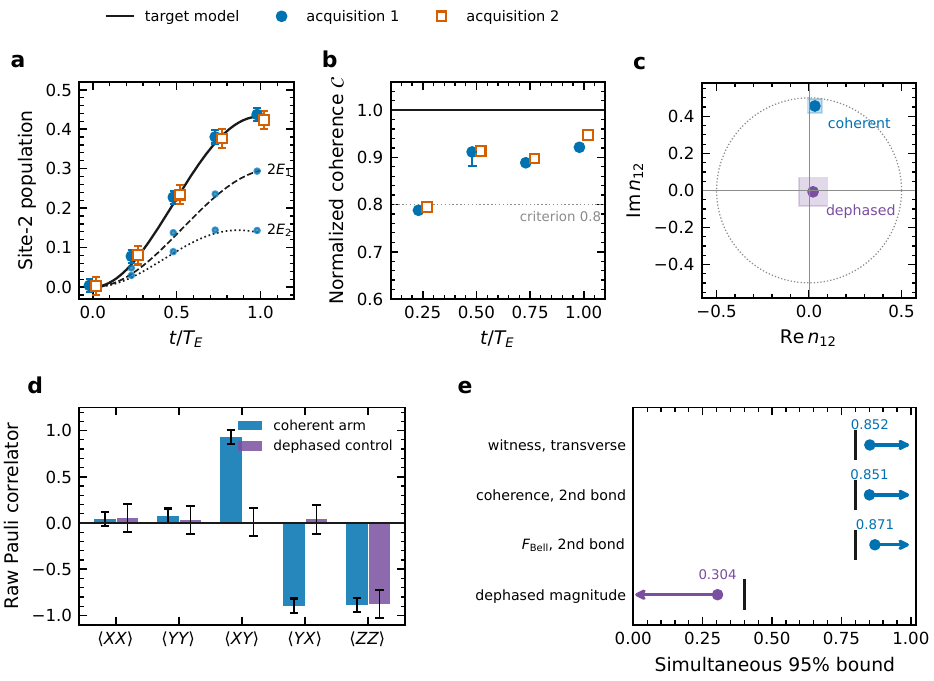}
\caption{\textbf{Coherence tests.} \textbf{a}, Transverse transfer: site-2 population from two acquisitions (markers, simultaneous 95\% intervals) and the target model (solid). Dashed and dotted curves with small markers resolve the two flavours of site 2. \textbf{b}, Normalized transverse coherence $\mathcal C$; the bar at half transfer marks the bootstrap lower bound 0.882. The ratio is undefined at $t=0$ and is not plotted there. \textbf{c}, Second bond at half transfer: raw intersite coherence of the coherent arm and of a matched dephased control, with simultaneous 95\% boxes; the dotted circle marks the largest attainable $|n_{12}|$ of 1/2. \textbf{d}, Raw Pauli correlators of the two arms. \textbf{e}, Prespecified decision quantities: lower bounds for the coherent tests (arrows to the right) and an upper bound for the dephased magnitude (arrow to the left); vertical ticks mark the criteria 0.8 and 0.4.}\label{fig:coherence}
\end{figure}

The transverse experiment prepares a richer logical state. Its off-diagonal site block $C_{12}$ is a complex $2\times2$ matrix in flavour space, and we summarize it by the normalized coherence $\mathcal C=\sigma_1(C_{12})/\sqrt{P_1P_2}$, where $P_i$ are the site populations. Both acquisitions meet the raw site-population, conditional-flavour and midpoint-coherence criteria. At half transfer $\mathcal C=0.91143$ and 0.91295, and the bootstrap lower bound of the first acquisition is 0.88180 (Fig.~\ref{fig:coherence}b). The final site-2 population is 0.43713, against a target of 0.43241; because the flavours mix during transfer, the site population saturates well below one (Fig.~\ref{fig:coherence}a). Supplementary Note~\ref{SI-sec:transverse} shows all measured components of $C_{12}$.

A ratio such as $\mathcal C$ is awkward for finite-sample inference. In synthetic data generated from a positive-state proxy, the ratio bootstrap covers the true half-transfer value in 99.2\% of 2,000 datasets, but only in 90.75\% at zero time, where coherence and one site population are both small and the mean estimate overshoots the truth by 0.08481. We therefore added an independent test based on a linear witness $w=2\Re(u^\dagger C_{12}v)$ with unit vectors fixed before the new acquisition. Since $w\le2\|C_{12}\|_2$, a lower bound on $w$ and an upper bound on $2\sqrt{P_1P_2}$ give a conservative lower bound on $\mathcal C$. The 36,864 shots yield $w=0.778834$ with Hoeffding radius 0.042435~\cite{Hoeffding1963} and a normalized lower bound of 0.852355, above the 0.8 criterion (Fig.~\ref{fig:coherence}e).

\subsection*{Coherent transfer and dephasing on a second bond}
A second bond provides a more demanding test. For the second-neighbour bond $\mathbf R=(0,1,-1)$ the bare axial hopping is only $t_{AA}=0.0179525$~meV, and the long evolution window increases the bare projected error to about 0.21, while the dressed error remains near $2.5\times10^{-4}$. This bond was kept out of the primary experiment.

Two population acquisitions of 114,688 shots each compare the bare and dressed predictions directly. The first meets its calibration, contrast and absolute-accuracy criteria. In the repeat, calibration and contrast pass and the dressed point estimates lie below the 0.04 total-variation limit, while two of the simultaneous upper bounds reach 0.041028 and 0.044467. Joint-outcome calibrations taken before and after each acquisition differ by at most 0.005127 and 0.004639 (Supplementary Note~\ref{SI-sec:heldout}).

A phase-resolved acquisition on the same bond then compares the coherent program with a matched dephased arm, using ten Pauli observables and 14,848 shots (Fig.~\ref{fig:coherence}c,d). The coherent arm gives $\langle XY\rangle=0.930664$ and $\langle YX\rangle=-0.893555$. This protocol selects the Bell phase opposite to Eq.~\eqref{eq:bell}, with score $(1-\langle ZZ\rangle+\langle XY\rangle-\langle YX\rangle)/4=0.927979$ and simultaneous lower bound 0.870609. The normalized-coherence lower bound of the coherent arm is 0.850853, and the upper bound on the coherence magnitude of the dephased arm is 0.303691. The two arms have nearly equal odd-sector probabilities, 0.943848 and 0.939453, so the phase contrast is accompanied by similar measured sector retention. All six prespecified criteria are met (Fig.~\ref{fig:coherence}e).

Population data alone would have missed this distinction. A compiler may factor a target as $U=DV$ with $D$ diagonal and execute only $V$; every computational-basis probability is then unchanged for every input, while general coherences are lost. Two 147,456-shot acquisitions built this way meet all fourteen simultaneous population intervals against the full-model reference (Supplementary Note~\ref{SI-sec:popcontrols}). The coherence and witness measurements above are the tests that a phase-discarding compilation would fail.

\subsection*{Choosing the representation size}
\begin{figure}[t]\centering
\includegraphics[width=\textwidth]{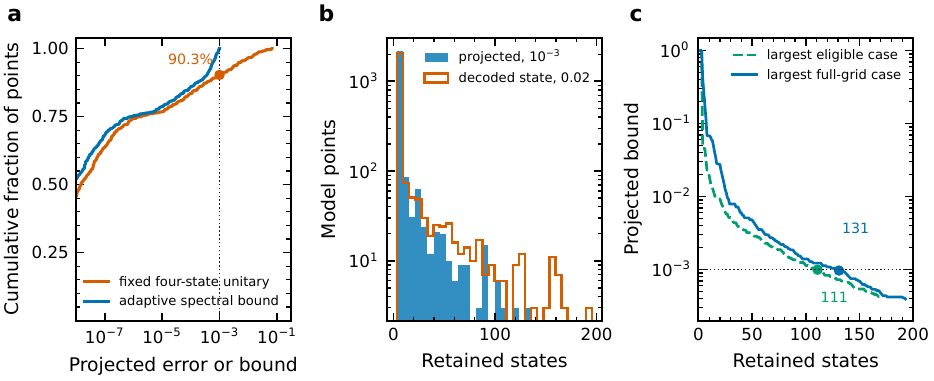}
\caption{\textbf{Accuracy and representation cost across the model class.} \textbf{a}, Cumulative distribution, over the 2,358 overlap-eligible points, of the projected error of the fixed four-state unitary and of the adaptive spectral bound of Eq.~\eqref{eq:adaptive} (16-state reference); the dotted line is the $10^{-3}$ tolerance. \textbf{b}, Number of retained states needed against the 495-state reference for the projected ($10^{-3}$) and decoded-state (0.02) tolerances, for all 2,520 points. \textbf{c}, Projected bound versus retained states for the most demanding overlap-eligible case and the most demanding case of the full grid; markers show the first dimensions that meet the tolerance.}\label{fig:resources}
\end{figure}

The primary experiment uses a favourable four-state reduction, and we asked how far that choice extends. Varying the parameters of the connected-bond model gives 2,520 points. Of the 2,358 points with adequate charge-transfer overlap, 2,130, or 90.33\%, meet a dressed projected-amplitude tolerance of $10^{-3}$ with four states (Fig.~\ref{fig:resources}a). The remainder is reproducible in the exact model: bare projected evolution is a contraction, and a fixed-size unitary cannot match every such contraction to arbitrary accuracy.

Static mixing explains much of this variation. Diagonalizing the retained and excluded bare blocks gives a first-order mixing matrix $(S_1)_{qa}=V_{qa}/(E_a-E_q)$ and $\theta_1=\|S_1\|_2$. For the 528 points that satisfy the overlap condition and have resolvable mixing, the ratio $\ell_{\max}/(2\theta_1)$ ranges from 0.92746 to 0.99977 with median 0.99800, inside the interval $[0.9,1.1]$ specified in advance. The observable also changes the verdict: of 634 grid points that fail the bare full-state criterion, 244 pass a population criterion of 0.02 for the axial input, so 38.49\% of these failures are failures only for a state-norm test. Equation~\eqref{eq:observable} accounts for this operator dependence.

A larger representation meets the same tolerance constructively. Select complete eigenvalue clusters $X_m$ and define $C_m=X_m^\dagger W$ and $G_m=C_m^\dagger C_m$. The projected spectral contraction and a normalized encoder satisfy
\begin{align}
K_m(t)&=C_m^\dagger e^{-iE_mt/\hbar}C_m, & \|W^\dagger UW-K_m\|_2&\le1-\lambda_{\min}(G_m),\nonumber\\
J_m&=C_mG_m^{-1/2}, & \|UW-X_me^{-iE_mt/\hbar}J_m\|_2&\le\sqrt{2-2\sqrt{\lambda_{\min}(G_m)}} .
\label{eq:adaptive}
\end{align}
Keeping whole clusters avoids splitting a near-degenerate level. Clusters are added in order of overlap weight per state until both bounds pass, so the resulting dimensions are constructive sufficient sizes. With this enlargement all 2,520 points reach the projected tolerance $10^{-3}$ and the decoded-state tolerance 0.02 (Fig.~\ref{fig:resources}b,c). The 16-state reference needs 4--13 states. Against the direct 495-state reference, the largest selected dimension is 131 for projected accuracy and 193 for decoded-state accuracy, and the projected maximum falls to 111 on the overlap-eligible subset. Eight logical qubits encode the largest selected space, and an explicit native decoder needs nine output qubits; state preparation and observable synthesis add to these counts.

When the selected subspace is computed instead of exact, its residual enters the budget. For an orthonormal trial matrix $\widehat Q$ with Hermitian generator $B$, residual $r_H=\|H\widehat Q-\widehat QB\|_2$ and $g=\lambda_{\min}(W^\dagger\widehat Q\widehat Q^\dagger W)$, a fixed-rank dressed bound is
\begin{equation}
2(1-\sqrt g)+1-g+2|t|r_H/\hbar .
\label{eq:residual}
\end{equation}
With outward rounding this becomes a test of the stored operator up to a specified time horizon. It certifies 17 of 24 additional native-model points at fixed rank, and enlarging to 4--11 states certifies both numerical tolerances at all 24 (Supplementary Note~\ref{SI-sec:adaptive}).

\section*{Discussion}
The measured coherences show what the processor did. Axial transfer on the primary bond gives two raw Bell-projector scores near 0.935, the transverse experiment resolves the complex intersite block and passes an independently fixed linear witness, and on the second bond a coherent arm is separated from a dephased control prepared under matched conditions. Each of these numbers is a property of executed circuits, with its shot count and confidence construction attached.

A complementary sampling study treats excitation centroids and fine splittings as separate targets when reconstructing relativistic multiplets from quantum-selected electronic spaces~\cite{YildirimQuantumSelection2026}.

The exact model supplies the other half of the argument. For the primary axial bond the bare encoding has a state-amplitude floor above 0.02, and a virtual correction of less than one percent to the hopping prevents a long-time projected error of 0.32. The device tests the dressed logical dynamics; the exact comparison decides how those dynamics correspond to a chosen material observable.

The calibration results explain how useful phase information can coexist with an unmet absolute-accuracy target. Both raw axial datasets resolve the sign of the coherence and give large Bell scores, while their simultaneous absolute intervals, which are much stricter, exceed $\pm0.05$ at some points. A calibration-aware analysis of the repeat meets the conditional criterion under explicit assumptions, and the second-bond population repeat resolves the bare/dressed contrast with two bounds just above their limit. More shots improve precision only within an adequate noise model. In prospective resampling tests a 286,720-shot design passes 394 of 400 and 390 of 400 stationary scenarios, but only 267 of 400 once a 0.003 preparation-mixture offset is added (Supplementary Note~\ref{SI-sec:stats}). The preparation model, more than the nominal shot total, decides whether such a forecast holds.

Several routes extend the experiment. A projected contraction can be implemented with an ancilla in place of a four-state unitary; the three-qubit dilation we verified locally requires 24 CX gates per time point and keeps the absolute weight of the ancilla-zero branch~\cite{Hu2020Dilation}. For inputs confined to the axial one-exciton sector, a two-CZ matchgate reproduces the relevant block with a continuous-window amplitude bound of $6.37\times10^{-5}$, a saving that comes from the restricted input domain. At the other extreme, a four-qubit circuit for the full 16-state bond needs 135 CZ gates in exact synthesis (Supplementary Note~\ref{SI-sec:alternatives}).

The adaptive construction gives a recipe for larger experiments: fix the error target, select a representation that satisfies a bound, and then add the cost of encoding and measurement. For the two-site observables studied here an exact classical reference is cheap, which is what makes the error budget complete. A many-body material will need independent constraints on the interactions and a reference strategy suited to the larger Hilbert space, and the bounds above give the quantities such a strategy must control.

\section*{Conclusion}
A dressed encoding lets a material-derived exciton bond be tested through phase-sensitive observables on a superconducting processor, and eight acquisitions resolve its axial and transverse coherent dynamics, including an independent bond and a dephasing control. Four logical states suffice at 90.33\% of the overlap-eligible points of the model class, and the adaptive construction gives sufficient dimensions under the stated spectral-selection rule at the rest. Where LiIrF$_6$ itself falls in that distribution depends on its intersite interactions, which a local spectral fit leaves undetermined (Supplementary Note~\ref{SI-sec:model}); measuring them is the next experiment this encoding calls for.

\section*{Methods}
\subsection*{Numerical model and effective operators}
The local basis contains one singlet and three triplet-derived states per ion. The two-ion construction starts from six native spinors and four electrons per ion and works in the combined eight-electron space. Basis transformations are applied consistently to Hamiltonians, states and observables, and the 16-state product operator and its four-state targets are compared with direct calculations in the 495-state space. Effective operators follow Eq.~\eqref{eq:dressed}. The second-order comparison takes its energy denominators from the diagonal of the product-basis Hamiltonian and is then Hermitian-symmetrized; the mixing diagnostic $\theta_1$ instead uses eigenvalues of the retained and excluded blocks. One- and ten-window reduction curves use 401 and 4,001 time points. For bond-wide comparisons, time is scaled to each hopping and capped at $t/\hbar=2000$~meV$^{-1}$. The model-class grid combines $U=2,2.5,3,3.5,4$~eV, $J_H=0.3,0.4,0.5,0.6$~eV, hopping multipliers $0.5,0.75,1,1.5,2,3$ and the 21 connected bonds. Eigenvalues within $10^{-6}$~meV are grouped into clusters, which are ordered by decreasing input-overlap weight per eigenstate, with the first spectral index breaking ties. The parameter points sample one model class; sampled maxima test the implementation, and the uniform statements follow from the algebraic bounds.

\subsection*{Circuit synthesis and execution}
All eight acquisitions used IBM Kingston. The primary axial, primary transverse, second-bond population and fixed-witness programs map the logical qubits to physical qubits $(21,36)$; the second-bond phase test uses $(118,129)$ for both of its arms. Effective propagators are synthesized directly in the native CZ, $R_z$ and $\sqrt{X}$ basis with Qiskit~\cite{Javadi2024Qiskit}, and Pauli frames around entangling gates implement randomized compiling~\cite{Wallman2016,Hashim2021}; ideal output distributions are checked after preparation and measurement rotations are included. The primary axial and transverse programs use five and nine Pauli settings. Their first acquisitions contain 442,368 and 278,528 shots and their independent repeats 442,368 and 155,648. The two second-bond population experiments contribute 114,688 shots each, the fixed transverse witness 36,864 and the phase test 14,848, for a total of 1,600,000 shots including calibration. Computational-basis preparations bracket the signal circuits to characterize assignment, using joint two-bit outcomes for each prepared state. Randomized circuits are pooled only within their measurement setting.

\subsection*{Observables and finite-shot uncertainty}
Raw Pauli moments are linear contrasts of measured counts, and the axial coherence follows Eq.~\eqref{eq:coherence}. For transverse data every entry of the complex block $C_{12}$ is a linear combination of Pauli moments, and $\mathcal C$ uses its largest singular value. Odd-sector conditioning divides an observable supported in that sector by its measured retention. Corrected and conditioned outputs are labelled separately from raw estimates. The primary axial criteria require simultaneous population and coherence agreement within 0.05, together with midpoint magnitude and sign tests. Transverse criteria require site-population agreement within 0.05, conditional-flavour agreement within 0.07 where the reference destination population exceeds 0.3, and a midpoint coherence lower bound of at least 0.8. Population envelopes use the more conservative of Clopper--Pearson and Hoeffding bounds~\cite{Clopper1934,Hoeffding1963}, and nonlinear ratios use the bootstrap~\cite{Efron1979}. The trajectory displays use a separate simultaneous 95\% family over 210 linear quantities with weighted Hoeffding intervals for independent bounded shots. Bell-projector and density-matrix interpretations require trusted local Pauli measurements of a common state.

\subsection*{Calibration-aware intervals}
The calibration-aware construction assumes a common assignment matrix for calibration and signal and a calibration-preparation error of at most $\delta=0.002$ in total variation. Clopper--Pearson calibration intervals widened by this allowance are combined with Hoeffding signal intervals in a simultaneous family of 145 events. Positivity and normalization define an outer set of pre-readout probabilities, and linear-fractional optimization with the Charnes--Cooper transformation gives the bounds on conditional observables~\cite{Charnes1962}. An independent ratio-bisection implementation agrees to $1.1\times10^{-13}$.

\subsection*{Code availability}
Analysis, replay, plotting and circuit-verification scripts have been uploaded with the data to the private Zenodo draft under reserved DOI \href{https://doi.org/10.5281/zenodo.23972323}{10.5281/zenodo.23972323}~\cite{Yildirim2026PaperAData}. Public access begins when the deposit is published. Qiskit, NumPy, SciPy and Matplotlib are obtained from their developers under their respective licences.

\section*{Data availability}
Measured counts, circuit payloads, model Hamiltonians, numerical source tables, source mappings and file hashes have been uploaded to a private Zenodo draft under reserved DOI \href{https://doi.org/10.5281/zenodo.23972323}{10.5281/zenodo.23972323}~\cite{Yildirim2026PaperAData}. Public access begins when the deposit is published.

\section*{Acknowledgements}
The principal electronic-structure calculations reported in this paper were fully performed using the EuroHPC Joint Undertaking (EuroHPC JU) supercomputer MareNostrum 5, hosted by the Barcelona Supercomputing Center (BSC). Access to MareNostrum 5 was provided through a national access call coordinated by the Scientific and Technological Research Council of Turkey (T\"UB\.{I}TAK). We gratefully acknowledge BSC, T\"UB\.{I}TAK, and the EuroHPC JU for providing access to these resources and supporting this research.

We acknowledge IBM Quantum services and the IBM Kingston processor used for the quantum-computing experiments. The conclusions are the authors' and should not be interpreted as an official IBM or IBM Quantum team position.

\section*{Competing interests}
The authors declare no competing interests.

\clearpage
\section*{Tool Use}
Numerical diagonalization and matrix analysis were used to construct the bond model, reduced operators and error bounds. Qiskit was used to synthesize, verify and run circuits on IBM Kingston; NumPy, SciPy and Matplotlib supported numerical analysis, uncertainty calculations and figure preparation. The electronic parent, computational procedures and software references are described in Methods and the Supplementary Information.

During the preparation of this manuscript, the corresponding author used generative artificial-intelligence assistants for language editing, text organization, reference checking. These tools were not a source of measured data, and no numerical result was accepted solely from their output. Numerical results were obtained using the computational methods and IBM hardware described in this work. The authors take full responsibility for the content.

\clearpage
\section*{Supplementary Information}
\setcounter{section}{0}\setcounter{figure}{0}\setcounter{table}{0}\setcounter{equation}{0}
\renewcommand{\theHsection}{S.\arabic{section}}
\renewcommand{\theHsubsection}{S.\arabic{section}.\arabic{subsection}}
\renewcommand{\theHfigure}{S.\arabic{figure}}
\renewcommand{\theHtable}{S.\arabic{table}}
\renewcommand{\theHequation}{S.\arabic{equation}}
\renewcommand{\thesection}{\arabic{section}}
\makeatletter
\renewcommand{\@seccntformat}[1]{\ifcsname format@#1\endcsname\csname format@#1\endcsname\else\csname the#1\endcsname\quad\fi}
\newcommand{\format@section}{Supplementary Note \thesection.\quad}
\makeatother
\renewcommand{\thefigure}{S\arabic{figure}}
\renewcommand{\thetable}{S\arabic{table}}
\renewcommand{\theequation}{S\arabic{equation}}

\section{Model, basis and target operators}\label{SI-sec:model}
The native two-ion Hamiltonian acts on eight electrons in twelve spinors, and its 495 determinants are the parent of both the 16-state product model and the direct reductions. Rebuilding the connected matrix in a common Fock basis reproduces it to $1.563039\times10^{-10}$~meV; transforming states and operators together leaves covariance and spectral residuals of $6.1008\times10^{-12}$ and $2.8990\times10^{-12}$~meV. For the primary bond, the four-state targets obtained from the 495-state reference and from the 16-state model agree to $2.802662\times10^{-5}$~meV per entry in the axial case and $4.737331\times10^{-10}$~meV in the transverse case, with sampled projected-dynamics differences of $7.754677\times10^{-4}$ and $8.042897\times10^{-4}$.

\paragraph{Local-basis enlargement.} Enlarging the local basis tests whether the four-state product is a sufficient description of the two-ion low-energy space. Table~\ref{SI-tab:basis} gives the four- and nine-state basis results at both Hund couplings. A wider test with four cluster fits, three Hubbard values, 21 bonds and two local basis sizes (504 records) gives a minimum Gram eigenvalue of 0.99808834 for the lowest-sixteen subspace of the nine-state product. The stricter criterion that the full overlap-selected 81-state space be complete passes in 222 of 252 nine-state bond cases. In those fitted models the four-state product reaches a selected-manifold Gram eigenvalue near 0.997620 but contains only seven of the lowest sixteen exact states, the missing ones being quintet-derived. The hardware target is the $U=3$~eV, $J_H=0.4$~eV model of the main text.

\paragraph{Scope of the local energies.} The 342.486927~meV axial energy belongs to the native model used to construct the executed bond. The separate PBEh025 family cluster has a first-triplet centroid of 481.326115~meV and a first excitation of 473.142384~meV. It is a different computed operator, not a measured peak used to certify this Hamiltonian. Their agreement is not assumed. A conditional local-fit test leaves additional anisotropic transition information out of the fit: the four fitted models have held-out anisotropy deviations of $-24.62$, $-48.88$, $-41.50$ and $-6.31$\%, despite total-susceptibility deviations of $-4.37$, $-3.20$, $-2.87$ and $-4.34$\%. The anisotropy errors are relative to a small anisotropic component, not to the total magnetic moment. Three diagnostic fits require negative crystal-field scales and hence change the original one-electron model. Matching scalar energies or averages does not determine a transition tensor.

\paragraph{Source-specific comparison.}
The PBEh025 value above is independently reproduced from the saved three excitation energies 473.142384299, 485.379555181 and 485.456406945~meV, whose common ground energy is $-18637.330989337584E_h$. Their mean is 481.326115475~meV. The native run \texttt{Li\_r70o\_AE2s\_\_uksPBEh025} uses 305 atomic-orbital functions, fourteen electrons in the ten ligand-$\sigma$/Ir-$5d$ orbitals, and all 38,760 spin-orbital determinants. Ir/F use ANO-RCC-VDZP, the nearest Li have explicit 2s functions, and an outer F/Li boundary and point-charge field enter the one-electron operator. The atomic mean-field SOC convention is $H_{\rm SO}=-2\sum_a i\,\mathrm{AMFI}_a\otimes s_a$ after transformation to the saved active orbitals. The spectrum is the PBEh025 family context in the companion quantum-selection study~\cite{YildirimQuantumSelection2026}; it is neither that study's VTZP hierarchy nor this paper's executed bond. The source spectrum, input hashes and the centroid recomputation are supplied in \texttt{A09\_spectrum\_identity.json}.

\paragraph{What a local fit constrains.} At fixed electron number $N$, fixed $J_H$ and fixed one-electron terms, the rotationally invariant Kanamori operator satisfies
\begin{equation}
\frac{\partial H_N}{\partial U}=\frac{N(N-1)}{2}\,I ,
\end{equation}
which equals $6I$ for $N=4$. Changing $U$ shifts every local level equally and leaves excitation energies and eigenvectors unchanged, whereas the charge curvature $E_3+E_5-2E_4$ changes by the same increment as $U$. Virtual charge processes therefore keep a $U$ dependence in the two-ion operator. In one fitted local model (Table~\ref{SI-tab:sU}) the local excitation energies change by less than $2\times10^{-11}$~meV across the scan, while the axial exchange changes by 15.75\%.

\begin{table}[h]\centering\small
\caption{Interaction sensitivity of one fitted local model. The fit target and all other parameters are fixed; these exchange values belong to the fitted model, not to the hardware target.}\label{SI-tab:sU}
\begin{tabular}{rrr}\toprule
$U$ (eV)&Charge curvature (eV)&Axial exchange magnitude (meV)\\\midrule
2.5&1.865729537&0.159694192\\
3.0&2.365729537&0.147457313\\
3.5&2.865729537&0.134541383\\\bottomrule
\end{tabular}
\end{table}

\paragraph{Logical bases and Pauli coefficients.} The axial basis is $(gg,gA,Ag,AA)$ and the transverse basis is $(1E_1,1E_2,2E_1,2E_2)$, both in bit order $q_1q_0$. Pauli strings and the main-text Hamiltonians use this same order. The tabulated matrices and coefficients below permit independent reconstruction of the logical generators. The transverse site degree of freedom is $q_1$ and flavour is $q_0$; these are different tensor factors from the two site-occupation qubits of the axial encoding.

\paragraph{Explicit logical Hamiltonians.} The native-reference targets, with a scalar shift removed, are
\begin{align}
H_A/\mathrm{meV}&=\begin{pmatrix}-342.4736325 & 0.0065043 & 0.0065043 & -0.2229290\\0.0065043 & 0.0272066 & 0.1629593 & 0.0037697\\0.0065043 & 0.1629593 & 0.0272066 & 0.0037697\\-0.2229290 & 0.0037697 & 0.0037697 & 342.4192194\end{pmatrix},\\
H_E/\mathrm{meV}&=\begin{pmatrix}0.0322720 & 0.0368403 & 0.0930210 & 0.0809499\\0.0368403 & -0.0322720 & 0.0809499 & 0.1493794\\0.0930210 & 0.0809499 & 0.0322720 & 0.0368403\\0.0809499 & 0.1493794 & 0.0368403 & -0.0322720\end{pmatrix}.
\end{align}
These displayed entries retain the seven decimal places in the saved protocol; executed propagators use their full stored precision. Table~\ref{SI-tab:pauli} gives the nonzero coefficients from $h_{\mu\nu}=\Tr[H(\sigma_\mu\otimes\sigma_\nu)]/4$. Removing the scalar does not change the evolution.
The four-state reduction within the 16-state reference is separately
\begin{equation}
H_d^{(16)}/\mathrm{meV}=\begin{pmatrix}-0.000187 & 0.006509 & 0.006509 & -0.222901\\0.006509 & 342.500652 & 0.162959 & 0.003768\\0.006509 & 0.162959 & 342.500652 & 0.003768\\-0.222901 & 0.003768 & 0.003768 & 684.892665\end{pmatrix}.
\end{equation}
Its difference from the bare submatrix is
\begin{equation}
(H_d^{(16)}-W^\dagger HW)/\mathrm{meV}=\begin{pmatrix}-0.000187 & -0.000097 & -0.000097 & -0.000376\\-0.000097 & -0.002003 & 0.001299 & -0.000093\\-0.000097 & 0.001299 & -0.002003 & -0.000093\\-0.000376 & -0.000093 & -0.000093 & -0.000501\end{pmatrix}.
\end{equation}
This matrix retains its original scalar convention. It is not numerically interchangeable, entry by entry, with the direct native-reference target above; the small difference after aligning scalar shifts is quantified in the text.
\begin{table}[p]\centering\small
\caption{Pauli coefficients (meV), with the left operator on $q_1$ and the right on $q_0$. Omitted terms vanish at saved precision.}\label{SI-tab:pauli}
\begin{tabular}{lrr}\toprule
String & Axial & Transverse\\\midrule
$I_1X_0$ & 0.005137000 & 0.036840300\\
$I_1Z_0$ & -171.223212975 & 0.032272000\\
$X_1I_0$ & 0.005137000 & 0.121200200\\
$X_1X_0$ & -0.029984850 & 0.080949900\\
$X_1Z_0$ & 0.001367300 & -0.028179200\\
$Y_1Y_0$ & 0.192944150 & 0.000000000\\
$Z_1I_0$ & -171.223212975 & 0.000000000\\
$Z_1X_0$ & 0.001367300 & 0.000000000\\
$Z_1Z_0$ & -0.027206575 & 0.000000000\\
\bottomrule\end{tabular}\end{table}
\paragraph{All-bond reduction test.} Table~\ref{SI-tab:bonds} evaluates all connected bonds with 401 times and the $Ag$ input for total variation. The window is $\min[\pi\hbar/(2|t_{AA}|),2000\hbar\,\mathrm{meV}^{-1}]$. The overlap statistic is $g_{\min}=\lambda_{\min}[(W^\dagger X)^\dagger(W^\dagger X)]$, not the unsquared singular value. Entries are sampled maxima; analytic bounds supply time-uniform statements.
\begin{table}[p]\centering\scriptsize
\caption{Complete 21-bond reduction. $\ell$: leakage amplitude; $\epsilon_b,\epsilon_d$: bare and dressed projected errors; TV: dressed population total variation; hopping in meV.}\label{SI-tab:bonds}
\begin{tabular}{rlrrrrrrr}\toprule
Bond & $\mathbf R$ & $|t_{AA}|$ & $T$ (fs) & $g_{\min}$ & $\ell$ & $\epsilon_b$ & $\epsilon_d$ & TV\\\midrule
0 & (0,0,1) & 0.161427 & 6404.9 & 0.9998229 & 0.0266 & 0.032 & 0.000354 & 0.00042\\
1 & (0,0,2) & 0.000685 & 1316423.9 & 1.0000000 & 8.59e-05 & 6.85e-05 & 3.76e-09 & 4.37e-09\\
2 & (0,1,-1) & 0.017952 & 57591.8 & 0.9998742 & 0.0224 & 0.211 & 0.000251 & 0.000264\\
3 & (0,1,0) & 0.161660 & 6395.6 & 0.9998232 & 0.0266 & 0.0319 & 0.000353 & 0.000419\\
4 & (0,2,-2) & 0.000548 & 1316423.9 & 1.0000000 & 4.18e-05 & 1.61e-05 & 1.46e-09 & 1.12e-09\\
5 & (0,2,-1) & 0.000032 & 1316423.9 & 1.0000000 & 1.44e-05 & 2.16e-06 & 9.41e-10 & 1.44e-10\\
6 & (0,2,0) & 0.000685 & 1316423.9 & 1.0000000 & 8.59e-05 & 6.86e-05 & 3.69e-09 & 4.37e-09\\
7 & (1,-2,1) & 0.000118 & 1316423.9 & 1.0000000 & 5.05e-05 & 2.5e-05 & 1.64e-09 & 1.3e-09\\
8 & (1,-1,-1) & 0.003198 & 323304.3 & 1.0000000 & 0.000175 & 7.04e-05 & 1.53e-08 & 1.66e-08\\
9 & (1,-1,0) & 0.017873 & 57847.5 & 0.9998743 & 0.0224 & 0.212 & 0.000251 & 0.000264\\
10 & (1,-1,1) & 0.003195 & 323611.2 & 1.0000000 & 0.000176 & 7.04e-05 & 1.54e-08 & 1.68e-08\\
11 & (1,0,-2) & 0.000032 & 1316423.9 & 1.0000000 & 1.44e-05 & 2.17e-06 & 1.25e-09 & 1.39e-10\\
12 & (1,0,-1) & 0.017976 & 57517.1 & 0.9998742 & 0.0224 & 0.211 & 0.000252 & 0.000263\\
13 & (1,0,0) & 0.161660 & 6395.6 & 0.9998230 & 0.0266 & 0.0319 & 0.000354 & 0.00042\\
14 & (1,1,-2) & 0.000117 & 1316423.9 & 1.0000000 & 5.02e-05 & 2.49e-05 & 1.37e-09 & 1.29e-09\\
15 & (1,1,-1) & 0.003198 & 323350.3 & 1.0000000 & 0.000176 & 7.06e-05 & 1.54e-08 & 1.62e-08\\
16 & (2,-2,0) & 0.000547 & 1316423.9 & 1.0000000 & 4.18e-05 & 1.61e-05 & 1.21e-09 & 9.23e-10\\
17 & (2,-1,-1) & 0.000119 & 1316423.9 & 1.0000000 & 5.05e-05 & 2.49e-05 & 1.37e-09 & 1.3e-09\\
18 & (2,-1,0) & 0.000032 & 1316423.9 & 1.0000000 & 1.44e-05 & 2.16e-06 & 1.33e-09 & 1.84e-10\\
19 & (2,0,-2) & 0.000548 & 1316423.9 & 1.0000000 & 4.18e-05 & 1.61e-05 & 9.18e-10 & 9.26e-10\\
20 & (2,0,0) & 0.000685 & 1316423.9 & 1.0000000 & 8.59e-05 & 6.86e-05 & 3.71e-09 & 4.36e-09\\
\bottomrule\end{tabular}\end{table}
\begin{table}[p]\centering\small
\caption{Local-basis audit at $U=3$ eV: worst values over 21 bonds. $n_{16}$ counts selected lowest-sixteen exact roots; $w_{16}$ is their smallest projection weight. $w_{\mathrm{sel}}$ and $g_{\mathrm{sel}}$ refer to the complete overlap-selected manifold.}\label{SI-tab:basis}
\begin{tabular}{rrrrrr}\toprule
$J_H$ (eV) & Local states & $n_{16}$ & $w_{16,\min}$ & $w_{\mathrm{sel},\min}$ & $g_{\mathrm{sel},\min}$\\\midrule
0.4 & 4 & 7 & 3.797e-13 & 0.999445 & 0.999295\\
0.4 & 9 & 16 & 0.9995 & 0.998867 & 0.997891\\
0.5 & 4 & 12 & 5.031e-12 & 0.757143 & 0.750909\\
0.5 & 9 & 16 & 0.9992 & 0.998030 & 0.996349\\
\bottomrule\end{tabular}\end{table}

\section{Reduction identities and observable bounds}\label{SI-sec:proofs}
\paragraph{Leakage floor and best bare unitary.} Let $W^\dagger W=I$, $P=WW^\dagger$, $Q=I-P$ and $U$ unitary. Applying $Q$ to $UW-WV$ removes the second term and proves $\|UW-WV\|_2\ge\|QUW\|_2$ for every logical unitary $V$. For the exact minimum, set $B=W^\dagger UW$; then
\begin{equation}
(UW-WV)^\dagger(UW-WV)=2I-B^\dagger V-V^\dagger B .
\end{equation}
For a right singular vector $x$ of $B$ belonging to $\sigma_{\min}(B)$, $\Re\langle Bx,Vx\rangle\le\sigma_{\min}(B)$, so the squared norm is at least $2(1-\sigma_{\min})$. The unitary $U_BV_B^\dagger$ built from $B=U_B\Sigma_BV_B^\dagger$ attains this value, and a unitary extension covers singular $B$. Since $B^\dagger B=I-(QUW)^\dagger QUW$, a numerically stable form of the minimum is $\ell\sqrt{2/(1+\sqrt{1-\ell^2})}$ with $\ell=\|QUW\|_2$. This is the polar (orthogonal Procrustes) solution~\cite{Schonemann1966}.

\paragraph{Dressed isometry.} For an invariant selected space $X$ with $W^\dagger X=U_A\Sigma V_A^\dagger$, the dressed isometry is $W_d=XV_AU_A^\dagger$. Since $\|W-XV\|_F^2=2m-2\Re\Tr(W^\dagger XV)$, the trace is largest when the singular directions are aligned, which gives the polar rotation; orthogonality and invariance then make the dressed evolution exact~\cite{Lowdin1950,desCloizeaux1960}. The second-order comparison uses
\begin{equation}
(H_{\rm SW})_{ab}=H_{ab}+\frac12\sum_{k\notin P}H_{ak}H_{kb}\left[\frac1{E_a-E_k}+\frac1{E_b-E_k}\right],
\end{equation}
with $E_a$, $E_k$ the diagonal entries of the product-basis matrix, followed by Hermitian symmetrization. Its largest difference from the polar matrix is $1.93\times10^{-6}$~meV. Across the 21 connected bonds the six first-neighbour-like cases have sampled leakage floors of 0.0224--0.0266; the three second-neighbour bonds with $t_{AA}\simeq0.018$~meV reach bare projected errors near 0.21 against a dressed error of about $2.5\times10^{-4}$.

\paragraph{Matched preparations and observables.} Let $J$ span an invariant subspace, $A=J^\dagger W$, $Q=I-JJ^\dagger$ and $R_\perp=QW$. Then $W=JA+R_\perp$ and $UW=JVA+UR_\perp$ with $V=e^{-iH_dt/\hbar}$. Expanding $W^\dagger U^\dagger OUW$ gives the matched term $A^\dagger V^\dagger J^\dagger OJVA$, two adjoint cross terms and an excluded term, and their norms give
\begin{equation}
\epsilon_O\le2\|A\|_2\|R_\perp\|_2\|J^\dagger OQ\|_2+\|R_\perp\|_2^2\|QOQ\|_2 ,
\end{equation}
uniformly in time and over all normalized logical inputs. If a computed reduced Hamiltonian differs from $H_d$ by $\epsilon_H$ in spectral norm, Duhamel's formula adds $2\|A\|_2^2\|J^\dagger OJ\|_2\min(2,|t|\epsilon_H/\hbar)$. A conditioned observable also has a denominator: if $d,d'>0$, $d\ge d_{\min}$, $|n-n'|\le\epsilon_n$, $|d-d'|\le\epsilon_d$ and $|n'|\le\|O\|d'$, then
\begin{equation}
\left|\frac nd-\frac{n'}{d'}\right|\le\frac{\epsilon_n+\|O\|\epsilon_d}{d_{\min}} ,
\end{equation}
so an unconditioned bound transfers to a parity-conditioned estimate only together with control of the retention. Related bounds for perturbed spectra and observable-dependent simulation errors appear in Refs.~\cite{Szabo2025Transitions,Granet2025Dilution}.

\begin{figure}[h]\centering
\includegraphics[width=.92\textwidth]{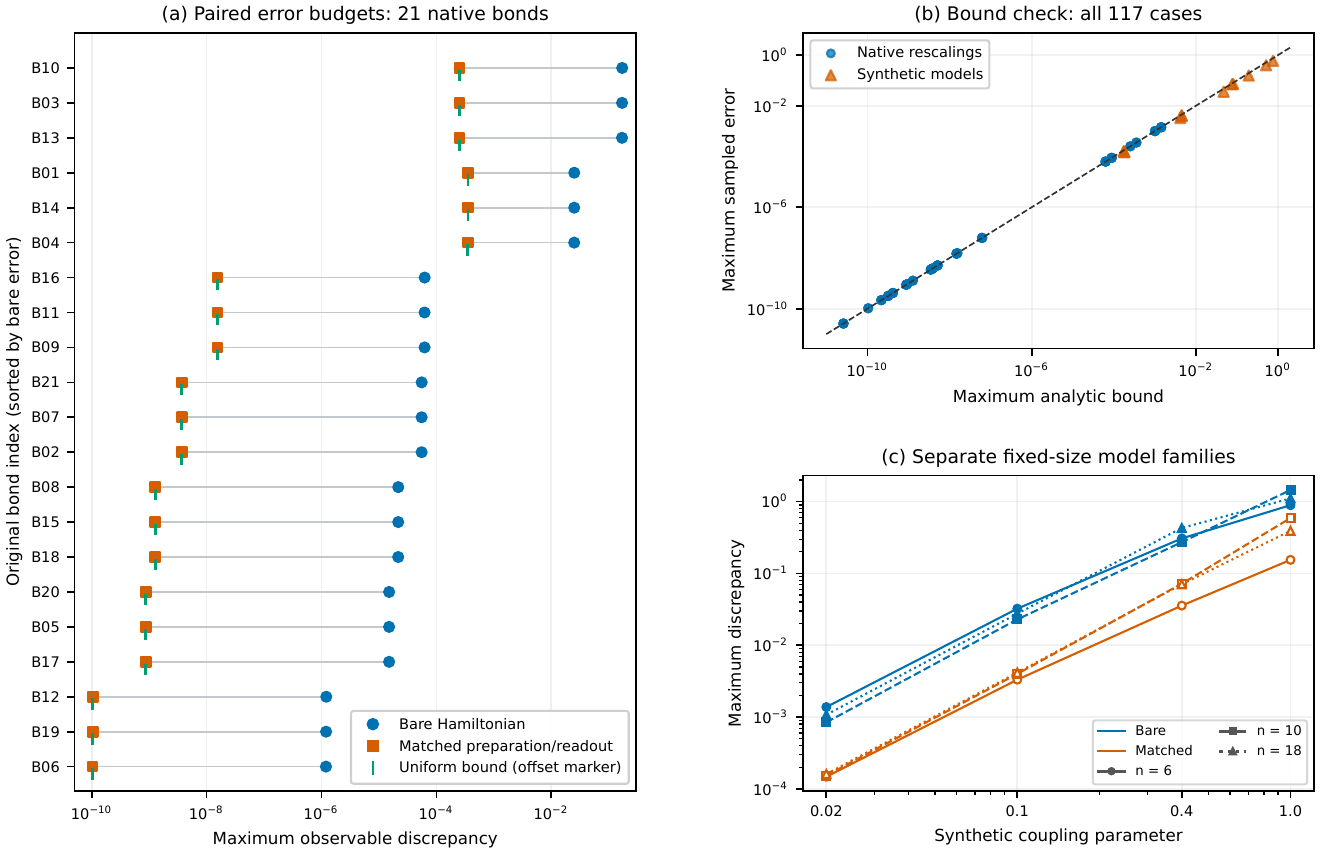}
\caption{Observable-bound checks on native bonds and synthetic models. Each point is the maximum error over seven observables and 61 times; the analytic bound is uniform in time. Synthetic curves connect cases of equal dimension as the coupling is varied.}\label{SI-fig:budget}
\end{figure}

\paragraph{Implementation checks.} The first test covers 21 native bonds at five rescalings and twelve synthetic Hermitian models of dimension 6, 10 and 18, in total 117 cases with seven observables and 61 times each (Fig.~\ref{SI-fig:budget}). The largest excess over the bound is $6.93\times10^{-15}$, and the minimax identity agrees with the explicit polar construction to $1.89\times10^{-15}$. A second test combines two encoded branches, 21 bonds and symmetric bit-assignment probabilities of 0, 0.02 and 0.10 (126 cases) with denominator-bounded conditional predictions; the largest error minus bound is $-4.14\times10^{-11}$.

\section{Circuits, synthesis and acquisitions}\label{SI-sec:circuits}
For the transverse generator the site qubit enters only through $I$ and $X$. In its $X$ basis the evolution is $|+\rangle\langle+|\otimes u_++|-\rangle\langle-|\otimes u_-$, a controlled unitary with Weyl coordinates $(a,0,0)$. At the four sampled nonzero times, the exact transverse and axial propagators need two and three CZ gates, respectively, as established by the invariants in main-text Table~\ref{tab:weyl-main}; special local or CNOT-equivalent times can need fewer. All 640 axial and 384 transverse nonzero-time circuits of the first payloads use these entangling counts. Measurement and preparation rotations are merged into the native single-qubit gates, and Pauli randomizations around the CZ gates preserve the ideal outcome distribution to $1.7\times10^{-15}$. The first four programs use 512 shots per compiled circuit and bracket the signal settings with eight copies of each of the four basis-state calibrations before and after. The first primary acquisitions executed on 21 October 2026 and both repeats on 24 October, separated by about 90.26~h (axial) and 92.30~h (transverse). Table~\ref{SI-tab:jobs} lists all eight acquisitions.

\begin{table}[h]\centering\small
\caption{The eight acquisitions of the phase-resolved study on IBM Kingston. Totals include calibration.}\label{SI-tab:jobs}
\begin{tabular}{lll}\toprule
Program & IBM job identifier & Shots\\\midrule
Axial, first &\texttt{dbbt661gv2cc73e94eag}&442,368\\
Axial, repeat &\texttt{dbehrs7sstms73fc430g}&442,368\\
Transverse, first &\texttt{dbbt6fpgv2cc73e94el0}&278,528\\
Transverse, repeat &\texttt{dbejlbom94ac738ruh30}&155,648\\
Second-bond populations 1 &\texttt{dbfr8r0m94ac738u6u6g}&114,688\\
Second-bond populations 2 &\texttt{dbg3fb9j4t8s73abhggg}&114,688\\
Fixed transverse witness &\texttt{dbfr96ngdj9s73bca670}&36,864\\
Second-bond phase test &\texttt{dbldpj58mdtc73d603ig}&14,848\\\midrule
Total &&1,600,000\\\bottomrule
\end{tabular}
\end{table}

\paragraph{Native gate census.} Table~\ref{SI-tab:gatecensus} lists all gate-count patterns in the two first signal payloads, including zero time and readout rotations, excluding their 64 calibration circuits each. Axial and transverse signal totals are 800 and 480; with calibration they are 864 and 544. These are executed circuits, distinct from the unexecuted alternatives in Note~\ref{SI-sec:alternatives}.
\begin{table}[p]\centering\small
\caption{Complete native-gate census of first signal payloads. $n$ is the number of circuits with the pattern; measurement operations are omitted.}\label{SI-tab:gatecensus}
\begin{tabular}{rrrrr@{\hspace{18pt}}rrrrr}\toprule
\multicolumn{5}{c}{Axial}&\multicolumn{5}{c}{Transverse}\\\cmidrule(lr){1-5}\cmidrule(lr){6-10}
CZ&$\sqrt X$&$R_z$&$X$&$n$&CZ&$\sqrt X$&$R_z$&$X$&$n$\\\midrule
0 & 0 & 0 & 1 & 32 & 0 & 0 & 0 & 0 & 32\\
0 & 2 & 3 & 0 & 64 & 0 & 1 & 1 & 0 & 16\\
0 & 2 & 4 & 0 & 64 & 0 & 1 & 2 & 0 & 16\\
3 & 13 & 14 & 0 & 5 & 0 & 2 & 2 & 0 & 8\\
3 & 13 & 15 & 0 & 30 & 0 & 2 & 3 & 0 & 16\\
3 & 13 & 16 & 0 & 56 & 0 & 2 & 4 & 0 & 8\\
3 & 13 & 17 & 0 & 41 & 2 & 10 & 13 & 0 & 44\\
3 & 13 & 18 & 0 & 27 & 2 & 10 & 13 & 1 & 54\\
3 & 13 & 19 & 0 & 1 & 2 & 10 & 14 & 0 & 88\\
3 & 14 & 13 & 0 & 10 & 2 & 10 & 14 & 1 & 110\\
3 & 14 & 14 & 0 & 50 & 2 & 10 & 15 & 0 & 41\\
3 & 14 & 15 & 0 & 128 & 2 & 10 & 15 & 1 & 47\\
3 & 14 & 16 & 0 & 160 &  &  &  &  & \\
3 & 14 & 17 & 0 & 93 &  &  &  &  & \\
3 & 14 & 18 & 0 & 35 &  &  &  &  & \\
3 & 14 & 19 & 0 & 4 &  &  &  &  & \\
\bottomrule\end{tabular}\end{table}
\begin{figure}[p]\centering
{\small\textbf{a} Axial, $t/T=0.25$}\\[1pt]\resizebox{\textwidth}{!}{\begin{quantikz}[row sep=0.25cm, column sep=0.14cm, font=\scriptsize]
\lstick{$q_1$} & \gate{R_z(-2.077)} & \gate{\sqrt{X}} & \gate{R_z(\pi/2)} & \gate{\sqrt{X}} &  & \ctrl{1} & \gate{R_z(-\pi)} & \gate{\sqrt{X}} & \gate{R_z(1.964)} &  & \ctrl{1} & \gate{R_z(-\pi)} & \gate{\sqrt{X}} & \gate{R_z(-\pi)} & \ctrl{1} & \gate{R_z(2.077)} & \gate{\sqrt{X}} & \gate{R_z(-\pi/2)} & \gate{\sqrt{X}} & \gate{R_z(\pi)} & \meter{} \\
\lstick{$q_0$} & \gate{R_z(0.506)} & \gate{\sqrt{X}} & \gate{R_z(-\pi/2)} & \gate{\sqrt{X}} & \gate{R_z(0.000)} & \control{} & \gate{\sqrt{X}} & \gate{R_z(0.392)} & \gate{\sqrt{X}} & \gate{R_z(-\pi)} & \control{} & \gate{\sqrt{X}} & \gate{R_z(0.131)} & \gate{\sqrt{X}} & \control{} & \gate{R_z(-0.000)} & \gate{\sqrt{X}} & \gate{R_z(-2.636)} & \gate{\sqrt{X}} & \gate{R_z(\pi/2)} & \meter{}
\end{quantikz}
}\\[5pt]
{\small\textbf{b} Axial, $t/T=0.75$}\\[1pt]\resizebox{\textwidth}{!}{\begin{quantikz}[row sep=0.25cm, column sep=0.14cm, font=\scriptsize]
\lstick{$q_1$} & \gate{R_z(0.053)} & \gate{\sqrt{X}} & \gate{R_z(\pi/2)} & \gate{\sqrt{X}} & \ctrl{1} & \gate{\sqrt{X}} & \gate{R_z(2.749)} &  &  & \ctrl{1} & \gate{R_z(-\pi)} & \gate{\sqrt{X}} &  &  & \ctrl{1} & \gate{R_z(-0.053)} & \gate{\sqrt{X}} & \gate{R_z(-\pi/2)} & \gate{\sqrt{X}} & \gate{R_z(-\pi)} & \meter{} \\
\lstick{$q_0$} & \gate{R_z(-1.624)} & \gate{\sqrt{X}} & \gate{R_z(\pi/2)} & \gate{\sqrt{X}} & \control{} & \gate{R_z(-\pi)} & \gate{\sqrt{X}} & \gate{R_z(1.964)} & \gate{\sqrt{X}} & \control{} & \gate{R_z(-\pi)} & \gate{\sqrt{X}} & \gate{R_z(2.748)} & \gate{\sqrt{X}} & \control{} & \gate{R_z(\pi)} & \gate{\sqrt{X}} & \gate{R_z(-1.624)} & \gate{\sqrt{X}} & \gate{R_z(\pi/2)} & \meter{}
\end{quantikz}
}\\[5pt]
{\small\textbf{c} Axial, $t/T=1.00$}\\[1pt]\resizebox{\textwidth}{!}{\begin{quantikz}[row sep=0.25cm, column sep=0.14cm, font=\scriptsize]
\lstick{$q_1$} & \gate{R_z(-3.113)} & \gate{\sqrt{X}} & \gate{R_z(-\pi/2)} &  &  & \ctrl{1} & \gate{R_z(-\pi)} & \gate{\sqrt{X}} & \gate{R_z(-\pi)} &  & \ctrl{1} & \gate{\sqrt{X}} & \gate{R_z(-\pi)} &  & \ctrl{1} & \gate{R_z(1.542)} & \gate{\sqrt{X}} & \gate{R_z(-\pi/2)} & \gate{\sqrt{X}} & \gate{R_z(\pi)} & \meter{} \\
\lstick{$q_0$} & \gate{R_z(0.934)} & \gate{\sqrt{X}} & \gate{R_z(-\pi/2)} & \gate{\sqrt{X}} & \gate{R_z(-\pi)} & \control{} & \gate{\sqrt{X}} & \gate{R_z(1.570)} & \gate{\sqrt{X}} & \gate{R_z(-\pi)} & \control{} & \gate{\sqrt{X}} & \gate{R_z(0.525)} & \gate{\sqrt{X}} & \control{} & \gate{R_z(-0.000)} & \gate{\sqrt{X}} & \gate{R_z(-2.208)} & \gate{\sqrt{X}} & \gate{R_z(\pi/2)} & \meter{}
\end{quantikz}
}\\[5pt]
{\small\textbf{d} Axial, $t/T=0.50$, second randomization}\\[1pt]\resizebox{\textwidth}{!}{\begin{quantikz}[row sep=0.25cm, column sep=0.14cm, font=\scriptsize]
\lstick{$q_1$} & \gate{R_z(-2.582)} & \gate{\sqrt{X}} & \gate{R_z(\pi/2)} & \gate{\sqrt{X}} & \ctrl{1} & \gate{R_z(-\pi)} & \gate{\sqrt{X}} & \gate{R_z(2.357)} &  & \ctrl{1} & \gate{R_z(-\pi)} & \gate{\sqrt{X}} &  &  & \ctrl{1} & \gate{R_z(2.582)} & \gate{\sqrt{X}} & \gate{R_z(-\pi/2)} & \gate{\sqrt{X}} & \gate{R_z(\pi)} & \meter{} \\
\lstick{$q_0$} & \gate{R_z(1.012)} & \gate{\sqrt{X}} & \gate{R_z(-\pi/2)} & \gate{\sqrt{X}} & \control{} & \gate{\sqrt{X}} & \gate{R_z(0.785)} & \gate{\sqrt{X}} & \gate{R_z(-\pi)} & \control{} & \gate{\sqrt{X}} & \gate{R_z(0.262)} & \gate{\sqrt{X}} & \gate{R_z(-\pi)} & \control{} & \gate{R_z(0.000)} & \gate{\sqrt{X}} & \gate{R_z(-1.012)} & \gate{\sqrt{X}} & \gate{R_z(-\pi/2)} & \meter{}
\end{quantikz}
}\\[5pt]
{\small\textbf{e} Transverse, $t/T=0.25$}\\[1pt]\resizebox{.85\textwidth}{!}{\begin{quantikz}[row sep=0.25cm, column sep=0.14cm, font=\scriptsize]
\lstick{$q_1$} & \gate{R_z(\pi/2)} & \gate{\sqrt{X}} & \gate{R_z(-1.461)} & \gate{\sqrt{X}} & \gate{R_z(0.000)} & \ctrl{1} & \gate{\sqrt{X}} & \gate{R_z(-2.772)} & \gate{\sqrt{X}} & \ctrl{1} & \gate{R_z(-\pi/2)} & \gate{\sqrt{X}} & \gate{R_z(-\pi/2)} & \gate{\sqrt{X}} & \gate{R_z(1.984)} & \meter{} \\
\lstick{$q_0$} & \gate{R_z(0.098)} & \gate{\sqrt{X}} & \gate{R_z(-1.901)} & \gate{\sqrt{X}} & \gate{R_z(-1.866)} & \control{} & \gate{X} & \gate{R_z(-1.697)} &  & \control{} & \gate{R_z(-0.295)} & \gate{\sqrt{X}} & \gate{R_z(-2.798)} & \gate{\sqrt{X}} & \gate{R_z(0.279)} & \meter{}
\end{quantikz}
}\\[5pt]
{\small\textbf{f} Transverse, $t/T=0.75$}\\[1pt]\resizebox{.85\textwidth}{!}{\begin{quantikz}[row sep=0.25cm, column sep=0.14cm, font=\scriptsize]
\lstick{$q_1$} & \gate{R_z(\pi/2)} & \gate{\sqrt{X}} & \gate{R_z(-1.232)} & \gate{\sqrt{X}} & \gate{R_z(-\pi)} & \ctrl{1} & \gate{R_z(-\pi)} & \gate{\sqrt{X}} & \gate{R_z(1.095)} & \gate{\sqrt{X}} & \ctrl{1} & \gate{R_z(-\pi/2)} & \gate{\sqrt{X}} & \gate{R_z(-\pi/2)} & \gate{\sqrt{X}} & \gate{R_z(-0.339)} & \meter{} \\
\lstick{$q_0$} & \gate{R_z(0.318)} & \gate{\sqrt{X}} & \gate{R_z(-1.857)} & \gate{\sqrt{X}} & \gate{R_z(-2.433)} & \control{} & \gate{X} & \gate{R_z(1.222)} &  &  & \control{} & \gate{R_z(-0.862)} & \gate{\sqrt{X}} & \gate{R_z(-2.717)} & \gate{\sqrt{X}} & \gate{R_z(0.816)} & \meter{}
\end{quantikz}
}\\[5pt]
{\small\textbf{g} Transverse, $t/T=1.00$}\\[1pt]\resizebox{.85\textwidth}{!}{\begin{quantikz}[row sep=0.25cm, column sep=0.14cm, font=\scriptsize]
\lstick{$q_1$} & \gate{R_z(\pi/2)} & \gate{\sqrt{X}} & \gate{R_z(-1.025)} & \gate{\sqrt{X}} & \gate{R_z(-\pi)} & \ctrl{1} & \gate{\sqrt{X}} & \gate{R_z(-1.439)} & \gate{\sqrt{X}} & \ctrl{1} & \gate{R_z(-\pi/2)} & \gate{\sqrt{X}} & \gate{R_z(-\pi/2)} & \gate{\sqrt{X}} & \gate{R_z(-0.023)} & \meter{} \\
\lstick{$q_0$} & \gate{R_z(0.456)} & \gate{\sqrt{X}} & \gate{R_z(-1.807)} & \gate{\sqrt{X}} & \gate{R_z(0.445)} & \control{} & \gate{R_z(2.005)} &  &  & \control{} & \gate{R_z(1.126)} & \gate{\sqrt{X}} & \gate{R_z(-0.510)} & \gate{\sqrt{X}} & \gate{R_z(-2.071)} & \meter{}
\end{quantikz}
}
\caption{Executed native circuits for the $XX$ setting. Half-transfer circuits for both channels appear in main-text Fig.~\ref{fig:circuits}; this figure completes the four time fractions and illustrates a second axial randomization. Preparation and readout rotations are included. Displayed angles are rounded; payloads retain full executed precision.}\label{SI-fig:sicircuits}
\end{figure}
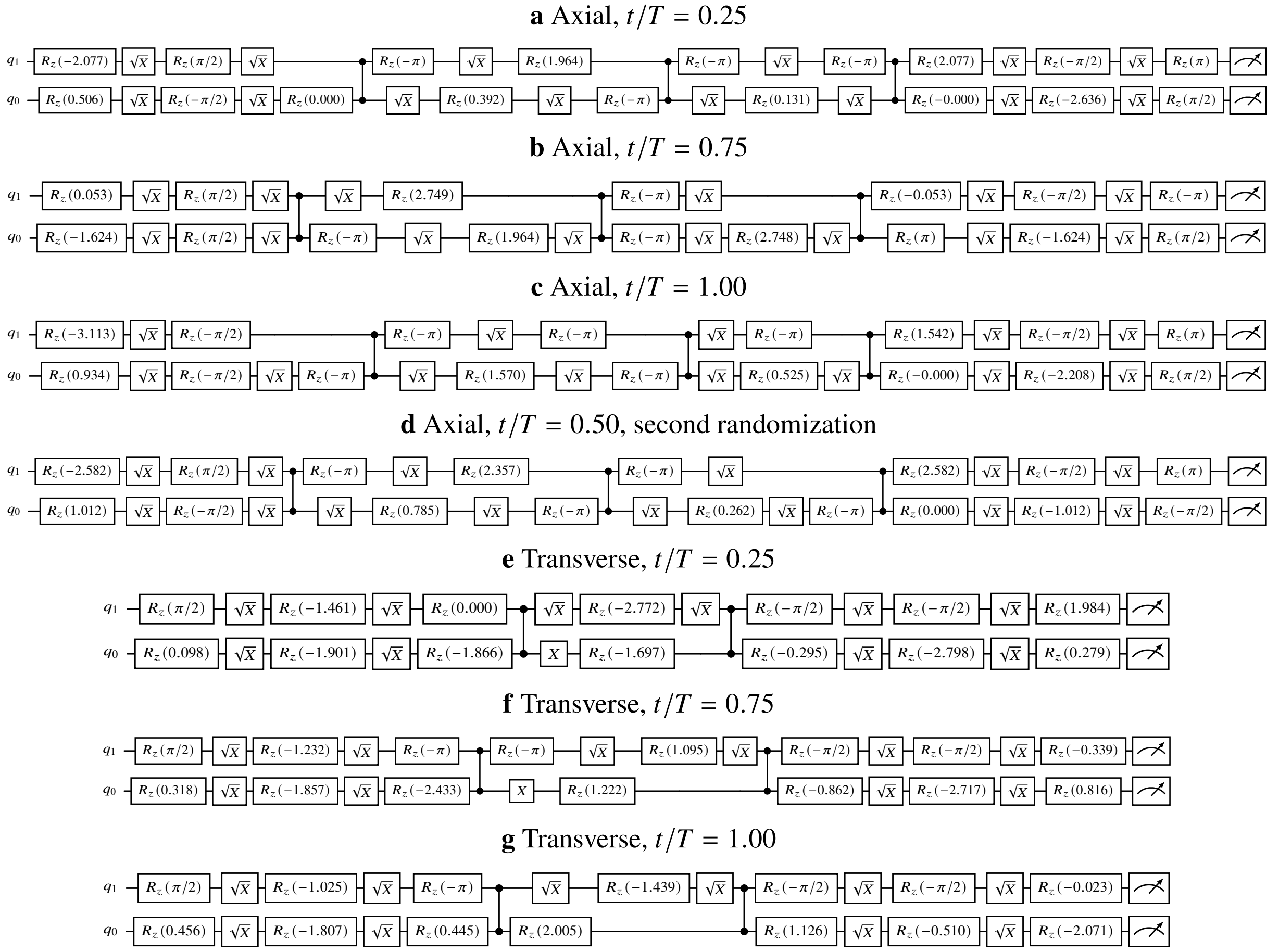
\paragraph{Calibration preparations.} Both channels prepare $|00\rangle$, $|01\rangle$, $|10\rangle$ and $|11\rangle$ by applying $X$ to the corresponding computational wires before measurement, both before and after the signal block. Calibration counts are pooled only within the prepared-state label. The full native census above includes every signal gate-count pattern; the circuit drawings are representative executed settings.

\begin{table}[tbp]\centering\scriptsize
\caption{Provider-recorded running and finished timestamps (UTC) for the eight main acquisitions. They are job-state timestamps, not pulse durations or calibration timestamps; created/queued times are not substituted for execution.}\label{SI-tab:acquisition-times}
\begin{tabular}{llll}\toprule
Acquisition & Running UTC & Finished UTC & Shots\\\midrule
XA original & 2026-10-21 02:51:37 & 2026-10-21 02:58:27 & 442,368\\
XE original & 2026-10-21 02:51:44 & 2026-10-21 02:56:28 & 278,528\\
XA repeat & 2026-10-24 21:06:57 & 2026-10-24 21:10:03 & 442,368\\
XE repeat & 2026-10-24 23:09:37 & 2026-10-24 23:10:47 & 155,648\\
Second-bond population 1 & 2026-10-26 22:05:24 & 2026-10-26 22:06:30 & 114,688\\
Second-bond population 2 & 2026-10-27 06:00:02 & 2026-10-27 06:02:22 & 114,688\\
Fixed transverse witness & 2026-10-26 22:05:28 & 2026-10-26 22:06:39 & 36,864\\
Second-bond phase & 2026-11-04 08:42:29 & 2026-11-04 08:42:46 & 14,848\\
\bottomrule\end{tabular}\end{table}
The timestamp records are bound to the job identifiers in the acquisition table by saved provider metrics. The contemporaneous assignment information comes from the computational-basis preparations measured within those programs; the matrices and their statistical assumptions are given in Supplementary Note~\ref{SI-sec:stats}. A complete per-job device-property history is not supplied, and no present-day calibration is used in its place. The separate three population-preserving compilation controls total 339,968 shots and are not included in the eight-acquisition total of 1,600,000.

\section{Estimators, confidence sets and calibration}\label{SI-sec:stats}
\paragraph{Parity and coherence.} For axial data the settings $ZZ,XX,YY,XY,YX$ give the populations and both quadratures of $n_{12}$. With $q=p_{01}+p_{10}$ and $\Pi=(I-ZZ)/2$, the identities $\Pi XX\Pi=(XX+YY)/2$ and $\Pi XY\Pi=(XY-YX)/2$ show that the supported coherence numerator divided by $q$ is the odd-sector coherence. Conditioning on detector outcomes and conditioning a reconstructed pre-readout distribution are different operations, and we keep them apart.

\paragraph{Acceptance families.} The raw axial family spreads an error probability of 0.05 over 32 decisions: four populations and two coherence components at five times, plus the midpoint magnitude and sign. It requires accuracy 0.05, a magnitude lower bound of 0.35 and a negative upper bound on the imaginary coherence. The transverse family has eleven decisions: site-population accuracy 0.05, conditional-flavour accuracy 0.07 where the reference destination population exceeds 0.3, and a midpoint coherence lower bound of at least 0.8. Population envelopes take the more conservative of the Clopper--Pearson and Hoeffding bounds~\cite{Clopper1934,Hoeffding1963}. The magnitude, sign and all transverse tests pass in both acquisitions; the raw axial absolute-accuracy tests exceed 0.05 at some points in both.

\paragraph{Assignment correction.} Corrected estimates use nonnegative least squares, $\min_{x\ge0}\|Ax-f\|_2$, followed by normalization~\cite{Bravyi2021Readout,Nation2021}. In the first axial acquisition the largest point errors after assignment correction, parity conditioning and both are 0.02841, 0.01375 and 0.00829; the combined correction in the repeat gives 0.00659. These are secondary endpoints and leave the raw criterion unchanged.

\begin{figure}[h]\centering
\includegraphics[width=.58\textwidth]{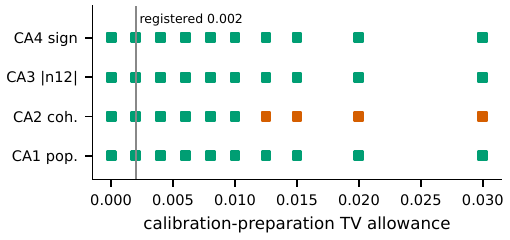}
\caption{Dependence of the four calibration-aware criteria of the axial repeat on the assumed preparation allowance $\delta$. CA1--CA4 denote population accuracy, coherence accuracy, coherence magnitude and coherence sign.}\label{SI-fig:slack}
\end{figure}

\paragraph{Calibration-aware outer intervals.} A common assignment matrix $A_{jk}=P(j|k)$ describes calibration and signal. For each prepared state and each of five binary events (four joint outcomes and even parity), calibration intervals widened by a preparation allowance $\delta$ give $l_{ek}\le A_{ek}\le u_{ek}$. Signal intervals $[f^-_{se},f^+_{se}]$ then imply
\begin{equation}
l_e\cdot p_s\le f^+_{se},\qquad u_e\cdot p_s\ge f^-_{se},\qquad p_s\ge0,\qquad\textstyle\sum_kp_{sk}=1 ,
\end{equation}
and the family of 145 events shares a total error probability of 0.05. Conditional populations and coherences are ratios of linear functions over this feasible set; Charnes--Cooper transformation~\cite{Charnes1962} and an independent ratio bisection agree to $1.1\times10^{-13}$. At $\delta=0.002$ the repeat meets its accuracy, magnitude and sign criteria (Table~\ref{SI-tab:ca}). Coherence accuracy holds up to $\delta=0.010$ and fails from 0.0125, while the other three criteria hold up to 0.03 (Fig.~\ref{SI-fig:slack}). The retrospective application of this same analysis to the first acquisition passes as well, with coherence accuracy up to $\delta=0.008$.

\begin{table}[h]\centering\small
\caption{Simultaneous calibration-aware outer intervals at preparation allowance $\delta=0.002$, conditioned on odd parity (family error 0.05). The 24 October repeat used the prespecified criterion; applying it to the first acquisition is retrospective. Reference populations and coherence follow the same dressed target.}\label{SI-tab:ca}
\begin{tabular}{lcccc}\toprule
Acquisition & $t/T_A$ & $P(gA)$ & $\Re n_{12}$ & $\Im n_{12}$\\\midrule
24 Oct (prospective) & 0.00 & $[0.00000,0.01687]$ & $[-0.02234,0.02807]$ & $[-0.02443,0.02600]$\\
 & 0.25 & $[0.12528,0.16676]$ & $[-0.01928,0.03283]$ & $[-0.38786,-0.31807]$\\
 & 0.50 & $[0.47433,0.51925]$ & $[-0.02406,0.02798]$ & $[-0.53147,-0.46062]$\\
 & 0.75 & $[0.82807,0.86976]$ & $[-0.02881,0.02319]$ & $[-0.38371,-0.31517]$\\
 & 1.00 & $[0.97686,1.00000]$ & $[-0.02674,0.02465]$ & $[-0.02581,0.02559]$\\
\midrule
21 Oct (retrospective) & 0.00 & $[0.00000,0.01701]$ & $[-0.02639,0.02466]$ & $[-0.02464,0.02643]$\\
 & 0.25 & $[0.12584,0.16762]$ & $[-0.02398,0.02865]$ & $[-0.38483,-0.31489]$\\
 & 0.50 & $[0.47319,0.51851]$ & $[-0.03046,0.02201]$ & $[-0.52666,-0.45604]$\\
 & 0.75 & $[0.83086,0.87283]$ & $[-0.03233,0.02038]$ & $[-0.38126,-0.31153]$\\
 & 1.00 & $[0.97525,1.00000]$ & $[-0.02831,0.02395]$ & $[-0.02374,0.02849]$\\
\midrule
\bottomrule\end{tabular}\end{table}

\begin{table}[h]\centering\scriptsize
\caption{Population, coherence-accuracy, magnitude and sign decisions, in that order, versus calibration-preparation allowance. P means pass and F fail. The first-acquisition reanalysis is retrospective.}
\begin{tabular}{lcccccccccc}\toprule
Acquisition & 0 & .002 & .004 & .006 & .008 & .010 & .0125 & .015 & .020 & .030\\\midrule
24 Oct & P/P/P/P & P/P/P/P & P/P/P/P & P/P/P/P & P/P/P/P & P/P/P/P & P/F/P/P & P/F/P/P & P/F/P/P & P/F/P/P\\
21 Oct & P/P/P/P & P/P/P/P & P/P/P/P & P/P/P/P & P/P/P/P & P/F/P/P & P/F/P/P & P/F/P/P & P/F/P/P & P/F/P/P\\
\bottomrule\end{tabular}\end{table}

\paragraph{Calibration stationarity.} Sixteen joint-outcome Fisher comparisons test calibration stability within the repeat; the smallest $p$-value is 0.074, above the family threshold of $6.25\times10^{-4}$. Joint outcomes matter here, since $(1/2,0,0,1/2)$ and $(0,1/2,1/2,0)$ have identical single-bit marginals. Between the two axial acquisitions, two of 25 time--setting cells differ after correction, the final $ZZ$ ($p=6.5\times10^{-6}$) and the midpoint $XY$ ($p=3.7\times10^{-4}$), with a largest outcome-probability shift of 0.0142. Table~\ref{SI-tab:homogeneity} gives all 25 comparisons. Failure to reject stationarity does not establish a time-independent detector outside the measured blocks.

\paragraph{Display intervals.} The trajectory plots contain 210 linear quantities: eight axial and thirteen transverse observables at five times in two acquisitions. For a linear contrast $L=\sum_s\overline{w_s(b)}$ of independent bounded shots, the radius
\begin{equation}
r_L=\left[\frac{\log(2/\alpha_*)}{2}\sum_s\frac{(\max_bw_s-\min_bw_s)^2}{N_s}\right]^{1/2},\qquad\alpha_*=0.05/210,
\end{equation}
gives simultaneous 95\% coverage by a union bound, including heterogeneous success probabilities across randomized shots. The Bell operator $[I-ZZ-XY+YX]/4$ has largest product-state overlap $1/2$, the square of the largest Schmidt coefficient, and convexity extends the bound to separable mixtures~\cite{GuhneToth2009}.

\paragraph{Shot allocation and preparation sensitivity.} Prospective design tests used 400 outer simulations and 5,000 bootstrap replicates each. At symmetric assignment error 0.02, a 512-shot-per-setting design passed 90 of 400 times and a 1,024-shot design 379 of 400 (one-sided 95\% lower bound 0.92528); at assignment error 0.08 the larger design passed 324 of 400, and an imposed dressed-population bias of 0.05 gave no passes in 400 trials. A later allocation of 286,720 shots passes 394 and 390 of 400 simulations in two empirical stationary scenarios (lower bounds 0.97061 and 0.95796), and 267 of 400 after a 0.003 preparation-mixture offset is added.

\paragraph{Full primary endpoints.} Tables~\ref{SI-tab:axraw}--\ref{SI-tab:secondary} retain raw estimates and secondary corrections. Raw population and coherence accuracy fail in both axial acquisitions; midpoint magnitude and sign pass. A corrected point estimate is not a calibration-aware confidence statement. S1 denotes nonnegative least-squares assignment correction, S2 measured odd-sector conditioning, and S1S2 their composition. The original recipe mentioned inverse-and-clip correction whereas the implementation used nonnegative least squares; these secondary results use the implemented estimator. The prospectively specified calibration-aware repeat is evaluated separately in Table~\ref{SI-tab:ca}.
\begin{table}[p]\centering\scriptsize
\caption{Raw axial endpoints. Last columns: population-envelope half-width and coherence-magnitude lower bound from the 32-decision family. Main-figure display intervals use a separate family.}\label{SI-tab:axraw}
\begin{tabular}{rrrrrrrrr}\toprule
$t/T_A$ & $p_{00}$ & $p_{01}$ & $p_{10}$ & $p_{11}$ & $\Re n_{12}$ & $\Im n_{12}$ & $w_p$ & $|n_{12}|_{\rm lo}$\\\midrule
\multicolumn{9}{l}{21 Oct 2026}\\
0.00 & 0.02576 & 0.00000 & 0.97223 & 0.00201 & 0.00687 & 0.00082 & 0.01478 & 0.00000\\
0.25 & 0.03259 & 0.13971 & 0.80487 & 0.02283 & 0.00977 & -0.32504 & 0.01478 & 0.31027\\
0.50 & 0.02673 & 0.47021 & 0.47766 & 0.02539 & 0.00385 & -0.46091 & 0.01478 & 0.44613\\
0.75 & 0.02734 & 0.80341 & 0.14087 & 0.02838 & 0.00226 & -0.32162 & 0.01478 & 0.30685\\
1.00 & 0.02240 & 0.94489 & 0.00690 & 0.02582 & 0.00571 & 0.00214 & 0.01478 & 0.00000\\
\multicolumn{9}{l}{24 Oct 2026}\\
0.00 & 0.03009 & 0.00000 & 0.96643 & 0.00348 & 0.00830 & 0.00073 & 0.01478 & 0.00000\\
0.25 & 0.03406 & 0.13983 & 0.79846 & 0.02765 & 0.01208 & -0.32742 & 0.01478 & 0.31265\\
0.50 & 0.02985 & 0.47198 & 0.47107 & 0.02710 & 0.00772 & -0.46353 & 0.01478 & 0.44876\\
0.75 & 0.02686 & 0.80658 & 0.14264 & 0.02393 & 0.00345 & -0.32486 & 0.01478 & 0.31008\\
1.00 & 0.02069 & 0.95575 & 0.00531 & 0.01825 & 0.00491 & -0.00012 & 0.01478 & 0.00000\\
\bottomrule\end{tabular}\end{table}
\begin{table}[p]\centering\scriptsize
\caption{Secondary axial point estimates. S1 and S2 address different ensembles; these values do not replace failed raw criteria. Finite calibration and preparation uncertainty are included in the separate calibration-aware intervals.}\label{SI-tab:secondary}
\begin{tabular}{lrrrrr}\toprule
Method & $t/T_A$ & $p_{01}$ & $p_{10}$ & $\Re n_{12}$ & $\Im n_{12}$\\\midrule
\multicolumn{6}{l}{21 Oct 2026, S1}\\
S1 & 0.00 & 0.00000 & 0.99879 & -0.00080 & 0.00087\\
S1 & 0.25 & 0.14346 & 0.82823 & 0.00224 & -0.33887\\
S1 & 0.50 & 0.48337 & 0.49142 & -0.00397 & -0.48041\\
S1 & 0.75 & 0.82603 & 0.14471 & -0.00561 & -0.33516\\
S1 & 1.00 & 0.97159 & 0.00685 & -0.00204 & 0.00226\\
\multicolumn{6}{l}{21 Oct 2026, S2}\\
S2 & 0.00 & 0.00000 & 1.00000 & 0.00706 & 0.00085\\
S2 & 0.25 & 0.14791 & 0.85209 & 0.01034 & -0.34411\\
S2 & 0.50 & 0.49607 & 0.50393 & 0.00406 & -0.48625\\
S2 & 0.75 & 0.85082 & 0.14918 & 0.00239 & -0.34061\\
S2 & 1.00 & 0.99275 & 0.00725 & 0.00600 & 0.00224\\
\multicolumn{6}{l}{21 Oct 2026, S1S2}\\
S1S2 & 0.00 & 0.00000 & 1.00000 & -0.00081 & 0.00087\\
S1S2 & 0.25 & 0.14764 & 0.85236 & 0.00231 & -0.34874\\
S1S2 & 0.50 & 0.49587 & 0.50413 & -0.00407 & -0.49283\\
S1S2 & 0.75 & 0.85093 & 0.14907 & -0.00578 & -0.34526\\
S1S2 & 1.00 & 0.99300 & 0.00700 & -0.00209 & 0.00231\\
\multicolumn{6}{l}{24 Oct 2026, S1}\\
S1 & 0.00 & 0.00000 & 0.99717 & 0.00279 & 0.00076\\
S1 & 0.25 & 0.14187 & 0.82375 & 0.00637 & -0.33972\\
S1 & 0.50 & 0.47980 & 0.48593 & 0.00184 & -0.48074\\
S1 & 0.75 & 0.82025 & 0.14707 & -0.00265 & -0.33695\\
S1 & 1.00 & 0.97210 & 0.00541 & -0.00099 & -0.00011\\
\multicolumn{6}{l}{24 Oct 2026, S2}\\
S2 & 0.00 & 0.00000 & 1.00000 & 0.00859 & 0.00076\\
S2 & 0.25 & 0.14903 & 0.85097 & 0.01288 & -0.34896\\
S2 & 0.50 & 0.50049 & 0.49951 & 0.00819 & -0.49152\\
S2 & 0.75 & 0.84973 & 0.15027 & 0.00363 & -0.34224\\
S2 & 1.00 & 0.99447 & 0.00553 & 0.00511 & -0.00013\\
\multicolumn{6}{l}{24 Oct 2026, S1S2}\\
S1S2 & 0.00 & 0.00000 & 1.00000 & 0.00279 & 0.00077\\
S1S2 & 0.25 & 0.14693 & 0.85307 & 0.00659 & -0.35182\\
S1S2 & 0.50 & 0.49682 & 0.50318 & 0.00191 & -0.49780\\
S1S2 & 0.75 & 0.84796 & 0.15204 & -0.00274 & -0.34833\\
S1S2 & 1.00 & 0.99447 & 0.00553 & -0.00101 & -0.00011\\
\bottomrule\end{tabular}\end{table}
\begin{table}[p]\centering\small
\caption{Raw transverse endpoints in the first acquisition. Zero-time coherence is a boundary diagnostic, not an acceptance endpoint. $P_2$ is destination population, $F_{E_2|2}$ conditional flavour, $[L,U]$ the simultaneous envelope for $P_2$.}\label{SI-tab:xeraw}
\begin{tabular}{rrrrrrr}\toprule
$t/T_E$ & $P_2$ & $[L,U]$ & Target & $F_{E_2|2}$ & Target & $\mathcal C$\\\midrule
0.00 & 0.0042 & [$-$0.0095, 0.0178] & 0.0000 & -- & -- & 0.2468 \\
0.25 & 0.0782 & [0.0646, 0.0918] & 0.0669 & -- & -- & 0.7878 \\
0.50 & 0.2277 & [0.2140, 0.2413] & 0.2245 & -- & -- & 0.9114 \\
0.75 & 0.3804 & [0.3667, 0.3940] & 0.3745 & 0.6202 & 0.6307 & 0.8885 \\
1.00 & 0.4371 & [0.4235, 0.4508] & 0.4324 & 0.6719 & 0.6832 & 0.9213 \\

\bottomrule\end{tabular}\end{table}
\begin{table}[p]\centering\small
\caption{Raw transverse repeat endpoints. The last column is the minimum eigenvalue of the linearly reconstructed density matrix; negativity describes finite-sample inversion. The midpoint positivity sensitivity is evaluated in Note~\ref{SI-sec:transverse}.}
\begin{tabular}{rrrrrrrr}\toprule
$t/T_E$ & $P_2$ & $[L,U]$ & Target & $F_{E_2|2}$ & Target & $\mathcal C$ & $\lambda_{\min}$\\\midrule
0.00 & 0.0027 & [$-$0.0166, 0.0220] & 0.0000 & -- & -- & 0.4753 & $-$0.01394 \\
0.25 & 0.0807 & [0.0614, 0.1000] & 0.0669 & -- & -- & 0.7952 & $-$0.00839 \\
0.50 & 0.2340 & [0.2147, 0.2533] & 0.2245 & -- & -- & 0.9129 & $-$0.00976 \\
0.75 & 0.3761 & [0.3568, 0.3954] & 0.3745 & 0.6186 & 0.6307 & 0.8974 & $-$0.00360 \\
1.00 & 0.4233 & [0.4041, 0.4426] & 0.4324 & 0.6670 & 0.6832 & 0.9474 & $-$0.00009 \\

\bottomrule\end{tabular}\end{table}
\paragraph{Pooled assignment matrices.} Rows denote measured outcomes and columns prepared states, each in order $00,01,10,11$:
\begin{align}A^{(21\,\mathrm{Oct})}&=\begin{pmatrix}0.9922 & 0.0072 & 0.0245 & 0.0001\\0.0022 & 0.9724 & 0.0000 & 0.0109\\0.0056 & 0.0001 & 0.9716 & 0.0066\\0.0000 & 0.0203 & 0.0039 & 0.9824\end{pmatrix},\\A^{(24\,\mathrm{Oct})}&=\begin{pmatrix}0.9961 & 0.0060 & 0.0273 & 0.0002\\0.0022 & 0.9830 & 0.0000 & 0.0144\\0.0016 & 0.0000 & 0.9691 & 0.0059\\0.0001 & 0.0110 & 0.0035 & 0.9795\end{pmatrix}.\end{align}
Column sums are one before rounding. Each matrix uses 32,768 calibration shots. Sixteen joint-outcome Fisher tests at family error 0.01 give smallest $p$ values 0.04966 and 0.07355. The constructed redistribution $(500,0,0,500)\rightarrow(0,500,500,0)$ gives joint-test $p=9.57\times10^{-188}$ but single-bit $p=1$, illustrating correlations missed by marginal readout checks.
\begin{table}[p]\centering\small
\caption{Between-acquisition Pearson homogeneity tests of pooled four-outcome distributions. Stars denote $p<0.05/25$. Last column: largest individual outcome-probability shift over the five settings.}\label{SI-tab:homogeneity}
\begin{tabular}{rrrrrrr}\toprule
$t/T_A$ & $ZZ$ & $XX$ & $YY$ & $XY$ & $YX$ & Max shift\\\midrule
0.00 & $0.00225$ & $0.104$ & $0.349$ & $0.415$ & $0.401$ & 0.00958\\
0.25 & $0.0368$ & $0.249$ & $0.384$ & $0.966$ & $0.0573$ & 0.01422\\
0.50 & $0.21$ & $0.196$ & $0.476$ & $0.000374^{*}$ & $0.172$ & 0.00983\\
0.75 & $0.0868$ & $0.665$ & $0.492$ & $0.0293$ & $0.0341$ & 0.01398\\
1.00 & $6.47e-06^{*}$ & $0.00982$ & $0.314$ & $0.324$ & $0.439$ & 0.01147\\
\bottomrule\end{tabular}\end{table}

\FloatBarrier
\section{Transverse complex coherence and the fixed witness}\label{SI-sec:transverse}
\begin{figure}[h]\centering
\includegraphics[width=\textwidth]{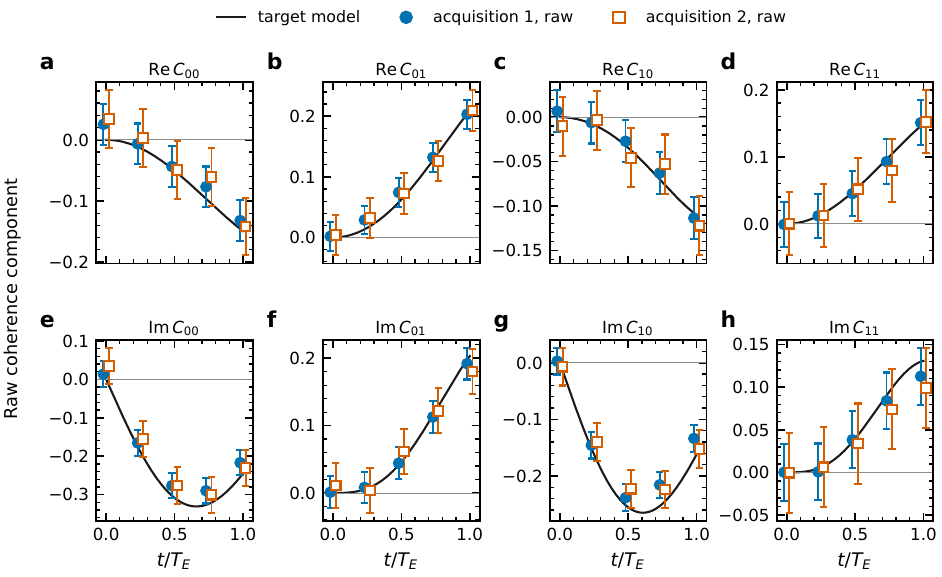}
\caption{Real (\textbf{a}--\textbf{d}) and imaginary (\textbf{e}--\textbf{h}) components of the intersite flavour-coherence block $C_{ab}$ in both acquisitions, with simultaneous 95\% intervals from the 210-quantity family. No assignment inversion or positive-state fit enters these components. Black curves: dressed target model.}\label{SI-fig:block}
\end{figure}

In the ordered site--flavour basis define $C_{ab}=\rho_{2+a,b}$. Raw counts give
\begin{align}
\Re C_{00}&=\tfrac12[P_{XZ}(00)-P_{XZ}(10)],&\Im C_{00}&=\tfrac12[P_{YZ}(00)-P_{YZ}(10)],\nonumber\\
\Re C_{11}&=\tfrac12[P_{XZ}(01)-P_{XZ}(11)],&\Im C_{11}&=\tfrac12[P_{YZ}(01)-P_{YZ}(11)],\nonumber\\
C_{01}&=\tfrac14[\langle XX\rangle+\langle YY\rangle+i(\langle YX\rangle-\langle XY\rangle)],&
C_{10}&=\tfrac14[\langle XX\rangle-\langle YY\rangle+i(\langle YX\rangle+\langle XY\rangle)],
\end{align}
identities that we verified as $4\times4$ operator equations. All eight real components are shown in Fig.~\ref{SI-fig:block}. The linear display differs at finite counts from the primary tomography, which averages compatible one-qubit marginals over three settings, by at most 0.0124512. Linear inversion gives small negative density-matrix eigenvalues between $-0.014$ and $-0.0001$; projecting onto positive trace-one matrices changes the repeat midpoint ratio from 0.912947 to 0.908493, and a 5,000-resample bootstrap at tail probability $0.05/11$ then gives a lower quantile of 0.869793.

\begin{figure}[h]\centering
\includegraphics[width=\textwidth]{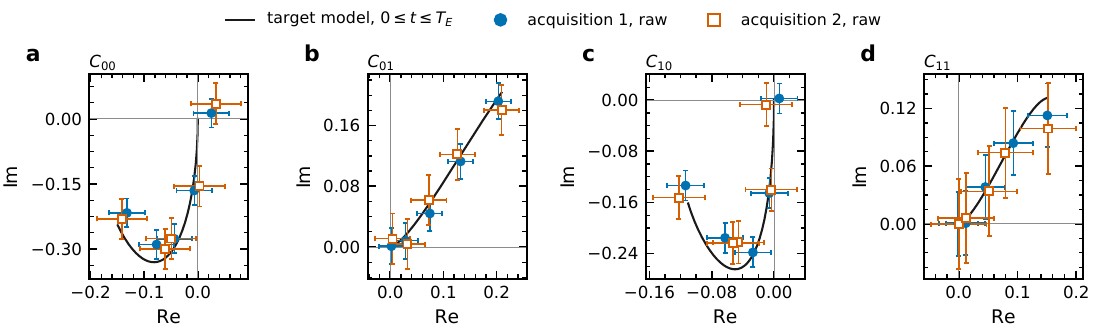}
\caption{Complex-plane trajectories of all four intersite flavour-coherence components $C_{ab}$ in both acquisitions. Horizontal and vertical bars use the simultaneous 95\% family of Fig.~\ref{SI-fig:block}; no assignment inversion or positive-state fit enters the displayed estimates. Black curves are the dressed target evolution from $t=0$ to $T_E$. The axial phase-space trajectory and odd-sector retention are shown in main-text Fig.~\ref{fig:axial}.}\label{SI-fig:traj}
\end{figure}

\paragraph{Bootstrap applicability.} A positive-state proxy fixed from a completed transverse acquisition generates 2,000 datasets at each of zero time and half transfer, with 2,000 inner bootstrap samples and tail $0.05/11$. At half transfer the proxy truth is 0.927480, the mean estimate 0.928273 and the lower-bound coverage 99.2\%. At zero time the truth is 0.340443, the mean estimate 0.425251 and the coverage 90.75\%.

\paragraph{Fixed witness.} Predetermined unit vectors $u,v$ define $w=2\Re(u^\dagger C_{12}v)$. With witness radius $r_w$ and simultaneous site-probability interval $[p_-,p_+]$,
\begin{equation}
\mathcal C\ge\frac{\max(0,w-r_w)}{2\max_{p\in[p_-,p_+]}\sqrt{p(1-p)}} .
\end{equation}
A retrospective witness chosen from one primary acquisition and evaluated on the other gives 0.815741; this is separate from the prospectively fixed witness. The dedicated witness program fixes its directions before collecting its 36,864 shots and gives 0.852355.

\section{Second-bond population and phase tests}\label{SI-sec:heldout}
\begin{figure}[h]\centering
\includegraphics[width=.9\textwidth]{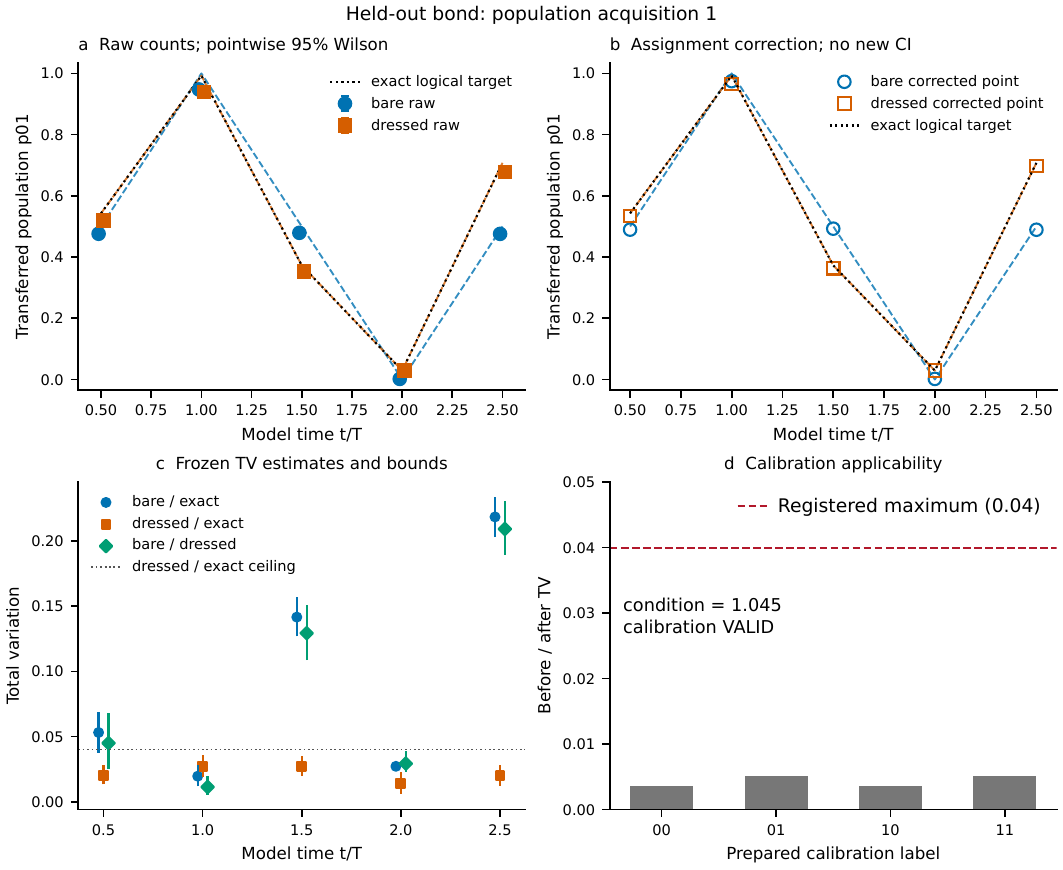}
\caption{First second-bond population acquisition: raw bare and dressed populations with pointwise Wilson bars, assignment-corrected values, simultaneous TV bounds and calibration checks. All four criteria are met.}\label{SI-fig:h1}
\end{figure}

\begin{figure}[h]\centering
\includegraphics[width=.9\textwidth]{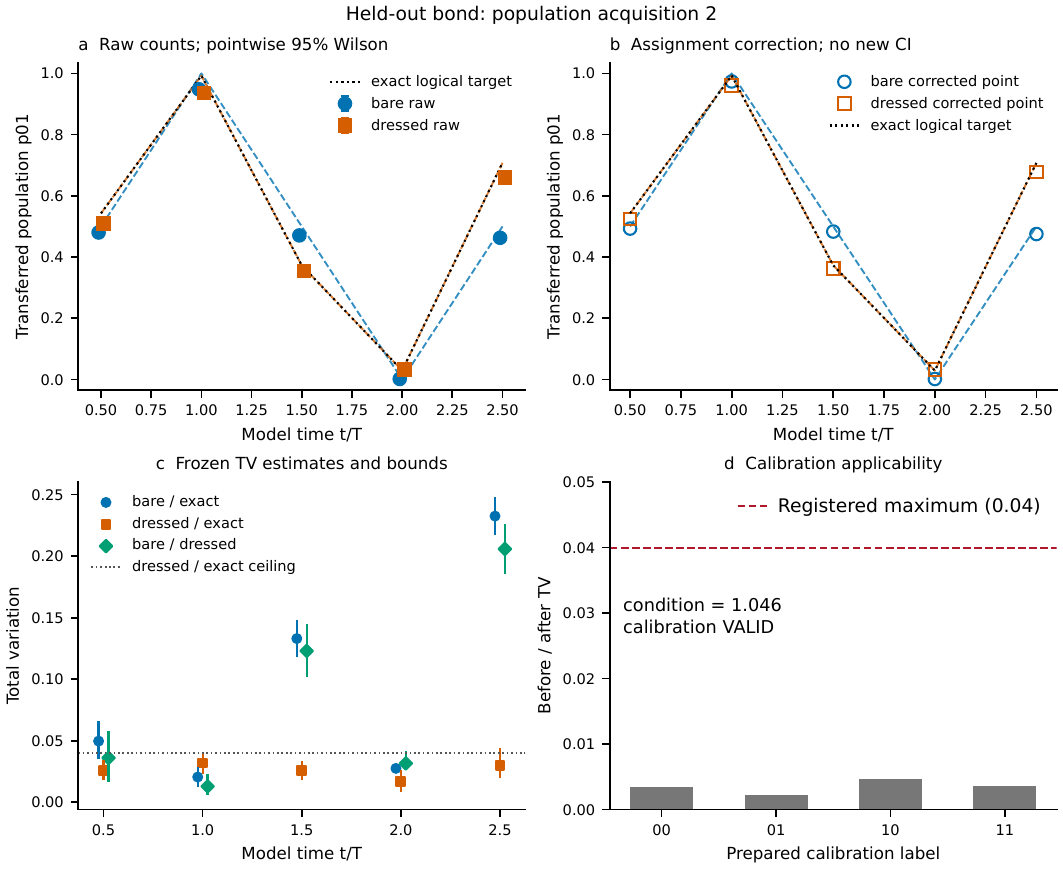}
\caption{Independent second-bond population repeat. Calibration and contrast criteria are met; two dressed TV upper bounds lie just above 0.04 while their point estimates lie below it.}\label{SI-fig:h2}
\end{figure}

The two 114,688-shot population programs compare bare and dressed dynamics on the second-neighbour bond (Figs.~\ref{SI-fig:h1} and \ref{SI-fig:h2}). The first meets all four criteria. In the repeat the two dressed TV upper bounds of 0.0410283 and 0.0444667 exceed 0.04 while all point estimates lie below it. Calibration condition numbers are 1.0455 and 1.0465, and the largest before/after joint-outcome TV changes are 0.005127 and 0.004639.


\paragraph{Exact confidence-set certificates.} A separate analysis uses sixteen calibration and forty science marginal intervals per acquisition. Under a fixed assignment matrix and independent, identically distributed pooled-setting shots, these constraints define an outer population polytope. For $\max c^Tz$ subject to $Az\le b$ and $0\le z\le1$, any nonnegative dual vector $y$ gives $c^Tz\le y^Tb+\sum_j\max[0,c_j-(A^Ty)_j]$. Directed binomial enclosures and exact rational arithmetic verify 590 stored dual certificates, with worst dressed-TV bounds of 0.04734 and 0.05270.

\begin{figure}[h]\centering
\includegraphics[width=.9\textwidth]{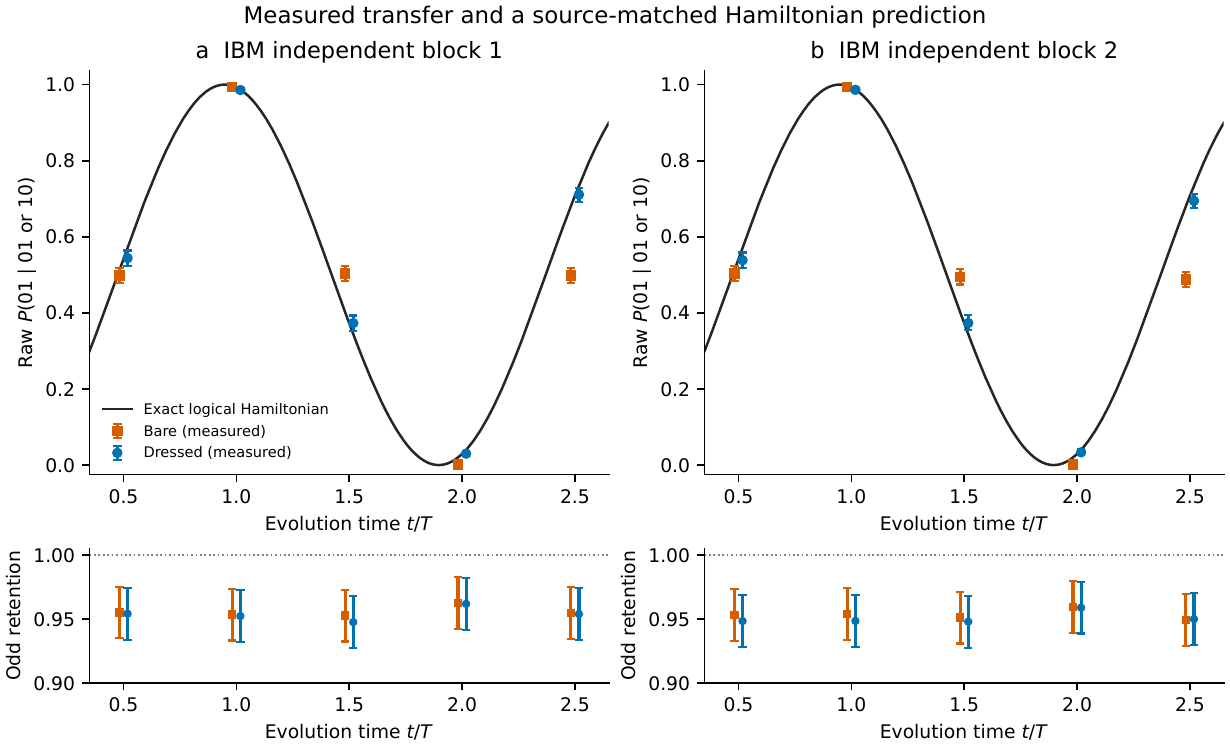}
\caption{Detector-conditioned transfer on the second bond, with the retained fraction shown explicitly.}\label{SI-fig:cond}
\end{figure}

\paragraph{Detector-conditioned transfer.} Conditioning on measured odd parity gives $r_{\rm det}=P(01)/[P(01)+P(10)]$. Simultaneous Bernstein inversion for heterogeneous independent shots gives largest absolute conditional errors of 0.02127 and 0.03232, with retention of 94.75--96.18\% and 94.79--95.89\% (Fig.~\ref{SI-fig:cond}). This secondary detector-conditioned analysis meets its conditional accuracy endpoint but does not meet the stronger bare/dressed contrast-interval analogue; it does not replace the unconditioned criteria.

\paragraph{Phase test.} The phase test uses a ten-observable family with 2,048 shots per coherent correlation and 512 per dephased correlation, pooling equal fixed dephasing strata. Its Bell operator is $[I-ZZ+XY-YX]/4$. The coherent score is 0.927979 with radius 0.057369 and the dephased score 0.461914 with radius 0.114738. Odd-sector probabilities are 0.943848 and 0.939453 with lower bounds 0.905602 and 0.862961. The normalized fixed-phase lower bound 0.850853 and the dephased magnitude upper bound 0.303691 meet the 0.8 and 0.4 thresholds; the assignment condition number 1.0380 and the calibration drift of 0.011719 in TV meet the applicability checks.

\section{Model-class generality and residual certification}\label{SI-sec:adaptive}
The 2,520-point grid consists of correlated variations within one native model class, and 2,358 points have adequate charge-transfer overlap. The fixed four-state dressed representation meets the projected tolerance $10^{-3}$ at 2,130 of them (90.33\%). Among 634 bare-state failures, 244 pass a population TV of 0.02 for the $Ag$ input (38.49\%).

\paragraph{Leading-order screen.} Where the overlap is adequate and $2\theta_1\ge0.005$, the 528 points have $\ell_{\max}/(2\theta_1)$ between 0.92746 and 0.99977, median 0.99800. A retrospective dressed-error screen $2\theta_1^2\le10^{-3}$ agrees with the sampled reference in 2,355 of 2,358 decisions, with no false accepts and three false rejects. The diagnostic uses the exact operator to define the retained and excluded blocks; its retrospective agreement is not an independently validated, reference-free classifier. On the 1,131 points with $\theta_1>10^{-4}$ the median measured-to-predicted ratio is 0.99774 and the mean absolute relative error 0.918\%.

\paragraph{Retrospective statistical screens.} On the same structured grid, a classifier trained on the other five failure-containing bonds reaches logistic AUC 0.999761 with three false accepts and three false rejects; a depth-three tree makes none on that split. A parameter-only logistic fit has AUC 0.995694. A two-translation-family stress yields AUC 0.996053 but nineteen false accepts and nine false rejects at its chosen threshold. These results distinguish ranking performance from accurate threshold decisions. They share lattice structure and exact-reference-derived targets and do not establish transfer to an independent material; the analytic and residual bounds above do not rely on the classifier.

\paragraph{Adaptive spectral construction.} Choose complete eigenvalue clusters $X_m$, let $C=X_m^\dagger W$ and $G=C^\dagger C\succ0$. The omitted contribution to $W^\dagger UW$ has norm at most $\|W^\dagger(I-X_mX_m^\dagger)W\|_2=1-\lambda_{\min}(G)$. The normalized encoder $J=CG^{-1/2}$ satisfies $J^\dagger J=I$ and $W^\dagger X_mJ=G^{1/2}$, so the original and decoded isometries differ by $\sqrt{2-2\sqrt{\lambda_{\min}(G)}}$, a bound preserved by the evolution. At fixed rank the static polar bound $2(1-\sqrt g)+1-g$ accepts 2,128 of the 2,358 points with no false accepts and two false rejects against the sampled criterion. Direct floating-point comparison shows a maximum apparent bound excess of $3.03\times10^{-9}$, which motivates retaining the residual and outward-rounding terms below rather than interpreting the floating-point screen itself as an arithmetic certificate.

\begin{figure}[h]\centering
\includegraphics[width=.92\textwidth]{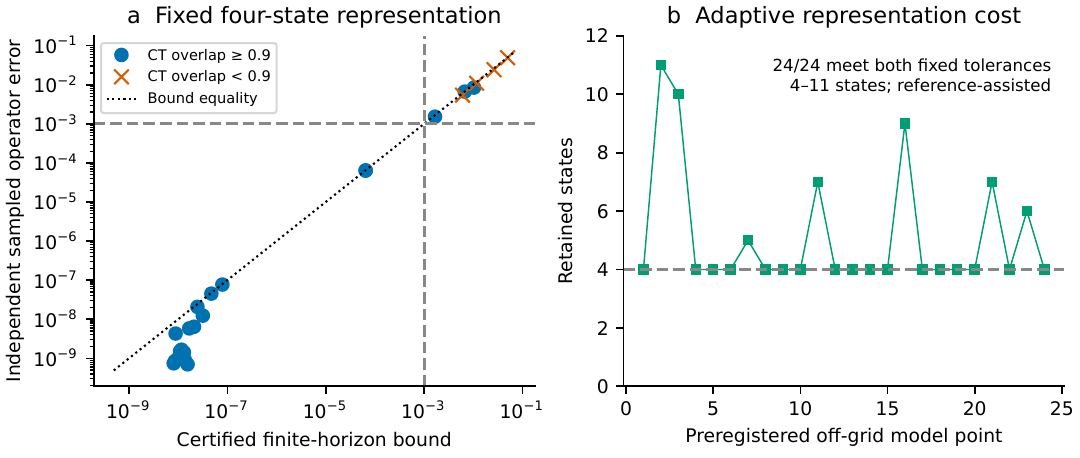}
\caption{Computed-subspace certification on synthetic and additional native-model points. \textbf{a}, Fixed-rank bound against an independent sampled operator error. \textbf{b}, Retained states needed to certify both tolerances at each of the 24 native points.}\label{SI-fig:cert}
\end{figure}

\paragraph{Computed subspaces.} For a computed orthonormal basis $\widehat Q$ with Hermitian generator $B$, Duhamel's formula gives $\|e^{-iH\tau}\widehat Q-\widehat Qe^{-iB\tau}\|_2\le|\tau|r_H$ with $r_H=\|H\widehat Q-\widehat QB\|_2$ and $\tau=t/\hbar$. The resulting bounds are $2(1-\sqrt g)+1-g+2|\tau|r_H$ at fixed rank, $1-g+2|\tau|r_H$ for a rectangular projected representation and $\sqrt{2-2\sqrt g}+|\tau|r_H$ for the decoded state. Exact rational products of the stored binary64 matrices, positivity tests and outward scalar enclosures account for the arithmetic. Fixed-rank certification succeeds at 17 of 24 additional native points, and enlarging to 4--11 states certifies both tolerances at all 24 (Fig.~\ref{SI-fig:cert}); four of these points carry a separate charge-transfer-overlap exclusion.

\section{Circuit alternatives and full-bond resources}\label{SI-sec:alternatives}
\begin{figure}[h]\centering
\includegraphics[width=.95\textwidth]{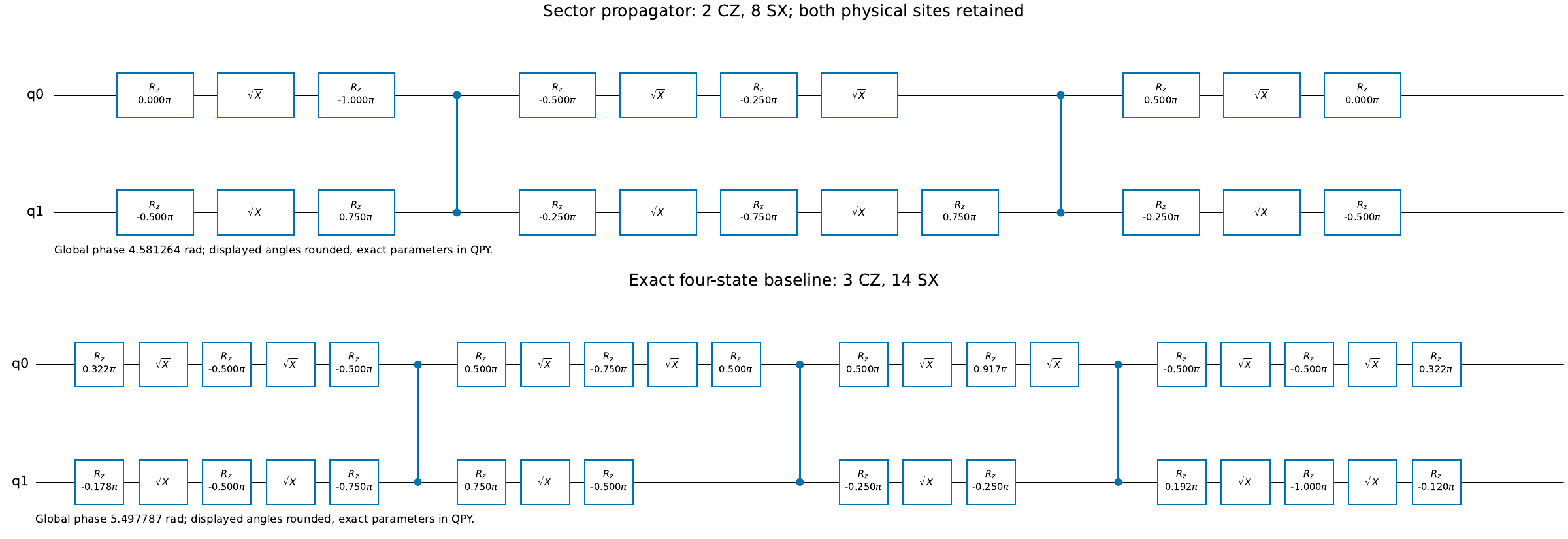}
\caption{Sector-specific two-CZ compilation and the complete-unitary baseline at half transfer, drawn from the saved native-basis candidates.}\label{SI-fig:sector}
\end{figure}

\paragraph{Sector circuit (not executed on hardware).} For the axial one-exciton subspace $Q_1=(|01\rangle,|10\rangle)$ let $h_1=Q_1^\dagger H_AQ_1$ and $a=\Tr(h_1)/2$. The extension
\begin{equation}
G(t)=e^{-iat/\hbar}(|00\rangle\langle00|+|11\rangle\langle11|)+Q_1e^{-ih_1t/\hbar}Q_1^\dagger
\end{equation}
has equal determinants in its even and odd blocks and is a two-entangler matchgate. It reproduces every coherent input in the one-exciton sector, with a continuous-window amplitude bound of $6.366\times10^{-5}$ and a largest sampled population discrepancy of $7.916\times10^{-7}$. The compiled candidate uses two CZ and eight $\sqrt X$ gates at depth 13, against three CZ and fourteen $\sqrt X$ at depth 22 for the complete unitary (Fig.~\ref{SI-fig:sector}).

\paragraph{Offline noise comparison.} A deterministic density-matrix proxy maps average gate infidelities to depolarizing channels and uses symmetric independent readout rates with separately assumed preparation errors. In the 23 October scenario the sector circuit has maximum population and coherence-component errors 0.05389 and 0.04797, compared with 0.05967 and 0.05179 for the complete unitary. Readout and preparation alone produce errors 0.03833 and 0.03809 in both circuits. Removing one entangler therefore does not by itself ensure raw accuracy. These conditional model expectations omit correlated, coherent, idle and drift errors and provide neither experimental intervals nor a validated prediction for another calibration.

\paragraph{One-ancilla dilation (not executed on hardware).} For a contraction $K$ on the four logical states,
\begin{equation}
\mathcal U_K=\begin{pmatrix}K&\sqrt{I-KK^\dagger}\\\sqrt{I-K^\dagger K}&-K^\dagger\end{pmatrix}
\end{equation}
acts on two register qubits and one ancilla and places $K$ in the ancilla-zero block~\cite{Hu2020Dilation}. Applied to $K_m(t)$ it carries the excluded probability in the other branch; the ancilla-zero event has probability $\|K_m(t)\psi\|^2$, and renormalizing it changes the prediction. Four constructions at window fractions $0.25,0.50,0.75,1$ have unitarity errors below $2.70\times10^{-15}$, synthesis discrepancies below $6.31\times10^{-14}$ and projected errors of $1.697\times10^{-4}$--$2.003\times10^{-4}$, under the uniform bound $4.694\times10^{-4}$; each flattened circuit has 24 CX gates at depth 49 before hardware routing. A separate dilation is synthesized at each time; this is not one autonomous, time-independent Hamiltonian realization. A one-qubit compression of the axial one-exciton sector reaches a state-amplitude error of $6.203\times10^{-5}$ over 401 times without entangling gates.

\paragraph{Full 16-state bond (not executed on hardware).} The complete bond requires four qubits. Exact compilation on a recorded target with physical wires $(151,138,152,150)$ has maxima of 135 CZ gates and depth 578 over the sampled family, with maximum full-operator discrepancy $7.52\times10^{-13}$. Lower-depth star-ansatz candidates with 36 and 48 CZ gates reach full-operator errors of 0.920792 and 0.676015, respectively; both exceed the 0.02 criterion. The best tested process infidelity is 0.127394.

\section{Population-preserving compilation}\label{SI-sec:popcontrols}
\begin{figure}[h]\centering
\includegraphics[width=\textwidth]{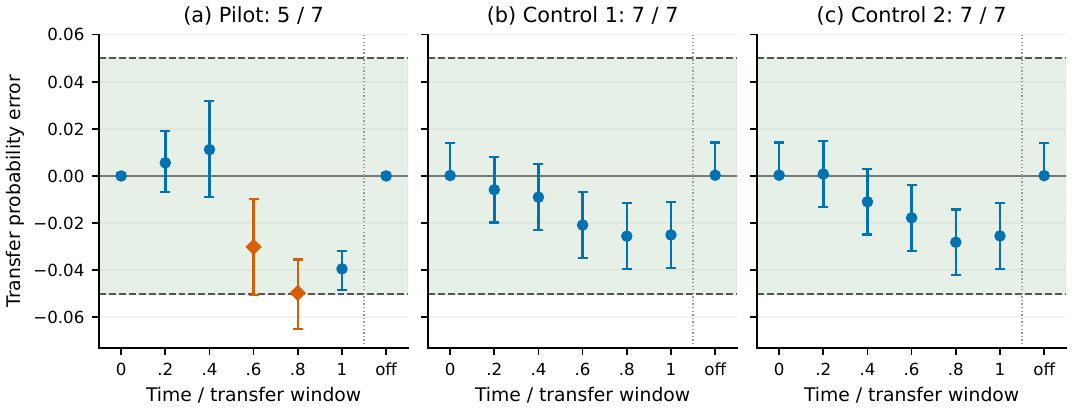}
\caption{Population errors of the pilot and the two population-preserving controls. The band marks the $\pm0.05$ tolerance; the last point in each panel is a no-evolution control.}\label{SI-fig:pop}
\end{figure}

For $U=DV$ with $D$ diagonal and unitary in the computational basis, every population projector $\Pi_z=|z\rangle\langle z|$ obeys
\begin{equation}
U^\dagger\Pi_zU=V^\dagger D^\dagger\Pi_zDV=V^\dagger\Pi_zV ,
\end{equation}
so $V$ reproduces the populations of $U$ for any input while phase-sensitive observables may change completely; in our implementation the population discrepancy is at most $5.99\times10^{-12}$ while the unitary discrepancy can approach 1.997. Each nonzero-time population circuit then needs two CZ gates instead of three. The target is the bare four-state block of the 16-state Hamiltonian at indices $(0,1,4,5)$, started in $|Ag\rangle$ over a 6395.63285~fs window and scored against the full 16-state probability of $|gA\rangle$ at $t/t_A^{(0)}=0,0.2,\ldots,1$, followed by a no-evolution control. A pilot with 4,096 science shots per endpoint meets five of its seven intervals; two follow-up controls, executed about 2.25 seconds apart in the same calibration period, with 114,688 science and 32,768 calibration shots each meet all fourteen (Table~\ref{SI-tab:pop}, Fig.~\ref{SI-fig:pop}). Reversible XOR decoding is applied to every outcome, no shot is discarded and no fitted assignment correction enters these endpoints. These three population-only acquisitions are additional to, and are not included in, the eight-acquisition total of 1,600,000 shots. Their total is 339,968 shots. The two successful controls do not test long-term calibration drift. The largest raw point errors are 0.0255803 and 0.0281438, and the contrast lower bounds are 0.946633 and 0.946084.

\begin{table}[h]\centering\small
\caption{Population-only acquisitions on IBM Kingston, physical qubits $(21,36)$. Totals include calibration.}\label{SI-tab:pop}
\begin{tabular}{llrr}\toprule
Program (execution date, UTC)&Job identifier&Shots&Intervals met\\\midrule
Pilot (19 Oct 2026)&\texttt{dbb197424mrc73865h3g}&45,056&5/7\\
Control 1 (20 Oct 2026)&\texttt{dbbg3loc77vs7384bg1g}&147,456&7/7\\
Control 2 (20 Oct 2026)&\texttt{dbbg3m9gv2cc73e8d5f0}&147,456&7/7\\\bottomrule
\end{tabular}
\end{table}

\clearpage

\clearpage
\bibliographystyle{naturemag}
\bibliography{references}
\end{document}

%% file: preamble.tex
\usepackage[a4paper,margin=2.4cm]{geometry}
\usepackage[T1]{fontenc}
\usepackage{amsmath}
\usepackage{newtxtext,newtxmath}
\usepackage{bm}
\usepackage{graphicx,booktabs,array,multirow}
\usepackage{xcolor}
\usepackage[font=small,labelfont=bf,labelsep=period,justification=justified]{caption}
\usepackage{subcaption}
\usepackage{tikz}
\usepackage{quantikz}
\usepackage[numbers,sort&compress,super]{natbib}
\usepackage{xr-hyper}
\usepackage[colorlinks=true,linkcolor=black,citecolor=black,urlcolor=black]{hyperref}
\usepackage{microtype}
\newcommand{\Tr}{\operatorname{Tr}}

\providecommand{\Re}{\operatorname{Re}}
\providecommand{\Im}{\operatorname{Im}}
\renewcommand{\Re}{\operatorname{Re}}
\renewcommand{\Im}{\operatorname{Im}}
\hypersetup{pdfauthor={Y\"ucel Y\i ld\i r\i m; \"Ozg\"ur Polat; Cengiz \c{S}en},pdfcreator={pdfLaTeX},pdfproducer={pdfTeX}}
\usepackage{pdflscape}
\usepackage{placeins}

%% file: circuits/xa_logical.tex
\begin{quantikz}[row sep=0.3cm, column sep=0.25cm, font=\footnotesize]
\lstick{$q_1$ (site 1)} & \gate{X}\gategroup[2,steps=1,style={dashed,rounded corners},label style={label position=above,yshift=-0.05cm}]{$|Ag\rangle$} & \gate[2]{U_{\mathrm{XA}}(t)=e^{-iH_{\mathrm{XA}}t/\hbar}} & \gate{B_{\sigma_1}}\gategroup[2,steps=1,style={dashed,rounded corners},label style={label position=above,yshift=-0.05cm}]{setting} & \meter{} \\
\lstick{$q_0$ (site 2)} & & & \gate{B_{\sigma_0}} & \meter{}
\end{quantikz}

%% file: circuits/xe_logical.tex
\begin{quantikz}[row sep=0.3cm, column sep=0.25cm, font=\footnotesize]
\lstick{$q_1$ (site)} & & \gate[2]{U_{\mathrm{XE}}(t)=e^{-iH_{\mathrm{XE}}t/\hbar}} & \gate{B_{\sigma_1}} & \meter{} \\
\lstick{$q_0$ (flavour)} & & & \gate{B_{\sigma_0}} & \meter{}
\end{quantikz}

%% file: circuits/XA_original_half_XX_tw0.tex
\begin{quantikz}[row sep=0.25cm, column sep=0.14cm, font=\scriptsize]
\lstick{$q_1$ (site 1)} & \gate{R_z(0.559)} & \gate{\sqrt{X}} & \gate{R_z(\pi/2)} & \gate{\sqrt{X}} & \ctrl{1} & \gate{\sqrt{X}} & \gate{R_z(-2.357)} &  &  & \ctrl{1} & \gate{R_z(-\pi)} & \gate{\sqrt{X}} &  & \ctrl{1} & \gate{R_z(0.559)} & \gate{\sqrt{X}} & \gate{R_z(-\pi/2)} & \gate{\sqrt{X}} & \gate{R_z(-0.000)} & \meter{} \\
\lstick{$q_0$ (site 2)} & \gate{R_z(1.012)} & \gate{\sqrt{X}} & \gate{R_z(\pi/2)} & \gate{\sqrt{X}} & \control{} & \gate{\sqrt{X}} & \gate{R_z(2.357)} & \gate{\sqrt{X}} & \gate{R_z(-\pi)} & \control{} & \gate{\sqrt{X}} & \gate{R_z(-2.879)} & \gate{\sqrt{X}} & \control{} & \gate{R_z(\pi)} & \gate{\sqrt{X}} & \gate{R_z(-2.130)} & \gate{\sqrt{X}} & \gate{R_z(\pi/2)} & \meter{}
\end{quantikz}

%% file: circuits/XE_original_half_XX_tw0.tex
\begin{quantikz}[row sep=0.25cm, column sep=0.14cm, font=\scriptsize]
\lstick{$q_1$ (site)} & \gate{R_z(\pi/2)} & \gate{\sqrt{X}} & \gate{R_z(-1.363)} & \gate{\sqrt{X}} & \gate{R_z(0.000)} & \ctrl{1} & \gate{R_z(-\pi)} & \gate{\sqrt{X}} & \gate{R_z(2.406)} & \gate{\sqrt{X}} & \ctrl{1} & \gate{R_z(-\pi/2)} & \gate{\sqrt{X}} & \gate{R_z(-\pi/2)} & \gate{\sqrt{X}} & \gate{R_z(2.410)} & \meter{} \\
\lstick{$q_0$ (flavour)} & \gate{R_z(0.202)} & \gate{\sqrt{X}} & \gate{R_z(-1.886)} & \gate{\sqrt{X}} & \gate{R_z(0.987)} & \control{} & \gate{R_z(1.815)} &  &  &  & \control{} & \gate{R_z(-2.557)} & \gate{\sqrt{X}} & \gate{R_z(-0.372)} & \gate{\sqrt{X}} & \gate{R_z(-2.590)} & \meter{}
\end{quantikz}